\RequirePackage{fix-cm}
\documentclass[pdftex,smallextended]{svjour3}       
\smartqed  
\usepackage{graphicx}
\usepackage{natbib}
\usepackage[breaklinks=true]{hyperref}
\usepackage[all]{hypcap}
\renewcommand{\url}[1]{\href{#1}{#1}}
\newcommand{\py}{\textsf{Py\-thon}}
\newcommand{\aw}{\textsf{Astro-WISE}}

\newcommand{\awe}{\texttt{AWE}}
\newcommand{\awenv}{\aw\ \textsf{Environment}}

\newcommand{\kids}{\textsf{KIDS}}
\newcommand{\readn}{\texttt{\href{http://doc.astro-wise.org/astro.main.ReadNoise.html}{Read\-Noise}}}
\newcommand{\gainl}{\texttt{\href{http://doc.astro-wise.org/astro.main.GainLinearity.html}{Gain\-Linearity}}}
\newcommand{\rawbi}{\texttt{\href{http://doc.astro-wise.org/astro.main.RawBiasFrame.html}{Raw\-Bias\-Frame}}}
\newcommand{\rawtw}{\texttt{\href{http://doc.astro-wise.org/astro.main.RawTwilightFlatFrame.html}{Raw\-Twilight\-Flat\-Frame}}}
\newcommand{\rawdo}{\texttt{\href{http://doc.astro-wise.org/astro.main.RawDomeFlatFrame.html}{Raw\-Dome\-Flat\-Frame}}}
\newcommand{\rawsc}{\texttt{\href{http://doc.astro-wise.org/astro.main.RawScienceFrame.html}{Raw\-Science\-Frame}}}
\newcommand{\biasf}{\texttt{\href{http://doc.astro-wise.org/astro.main.BiasFrame.html}{Bias\-Frame}}}
\newcommand{\darkc}{\texttt{\href{http://doc.astro-wise.org/astro.main.DarkCurrent.html}{Dark\-Current}}}
\newcommand{\twili}{\texttt{\href{http://doc.astro-wise.org/astro.main.TwilightFFlatFrame.html}{Twilight\-Flat\-Frame}}}
\newcommand{\domef}{\texttt{\href{http://doc.astro-wise.org/astro.main.DomeFlatFlatFrame.html}{Dome\-Flat\-Frame}}}
\newcommand{\night}{\texttt{\href{http://doc.astro-wise.org/astro.main.NightSkyFlatFrame.html}{Night\-Sky\-Flat\-Frame}}}
\newcommand{\maste}{\texttt{\href{http://doc.astro-wise.org/astro.main.MasterFlatFrame.html}{Master\-Flat\-Frame}}}
\newcommand{\fring}{\texttt{\href{http://doc.astro-wise.org/astro.main.FringeFrame.html}{Fringe\-Frame}}}
\newcommand{\coldp}{\texttt{\href{http://doc.astro-wise.org/astro.main.ColdPixelMap.html}{Cold\-Pixel\-Map}}}
\newcommand{\hotpi}{\texttt{\href{http://doc.astro-wise.org/astro.main.HotPixelMap.html}{Hot\-Pixel\-Map}}}
\newcommand{\reduc}{\texttt{\href{http://doc.astro-wise.org/astro.main.ReducedScienceFrame.html}{Reduced\-Science\-Frame}}}
\newcommand{\satur}{\texttt{\href{http://doc.astro-wise.org/astro.main.SaturatedPixelMap.html}{Saturated\-Pixel\-Map}}}
\newcommand{\satel}{\texttt{\href{http://doc.astro-wise.org/astro.main.SatelliteMap.html}{Satellite\-Map}}}
\newcommand{\cosmi}{\texttt{\href{http://doc.astro-wise.org/astro.main.CosmicMap.html}{Cosmic\-Map}}}
\newcommand{\astro}{\texttt{\href{http://doc.astro-wise.org/astro.main.AstrometricParameters.html}{Astro\-metric\-Parameters}}}
\newcommand{\gastr}{\texttt{\href{http://doc.astro-wise.org/astro.main.GAstrometric.html}{GAstro\-metric}}}
\newcommand{\phots}{\texttt{\href{http://doc.astro-wise.org/astro.main.PhotSrcCatalog.html}{Phot\-Src\-Catalog}}}
\newcommand{\photo}{\texttt{\href{http://doc.astro-wise.org/astro.main.PhotometricParameters.html}{Photo\-metric\-Parameters}}}
\newcommand{\illum}{\texttt{\href{http://doc.astro-wise.org/astro.main.IlluminationCorrectionFrame.html}{Illumination\-Correction\-Frame}}}
\newcommand{\regri}{\texttt{\href{http://doc.astro-wise.org/astro.main.RegriddedFrame.html}{Regridded\-Frame}}}
\newcommand{\coadd}{\texttt{\href{http://doc.astro-wise.org/astro.main.CoaddedRegriddedFrame.html}{Coadded\-Regridded\-Frame}}}
\newcommand{\weigh}{\texttt{\href{http://doc.astro-wise.org/astro.main.WeightFrame.html}{Weight\-Frame}}}
\newcommand{\sourc}{\texttt{\href{http://doc.astro-wise.org/astro.main.SourceList.html}{Source\-List}}}
\newcommand{\assoc}{\texttt{\href{http://doc.astro-wise.org/astro.main.AssociateList.html}{Associate\-List}}}
\newcommand{\combi}{\texttt{\href{http://doc.astro-wise.org/astro.main.CombinedList.html}{Combined\-List}}}
\newcommand{\proce}{\texttt{\href{http://doc.astro-wise.org/astro.main.ProcessTarget.html}{Process\-Target}}}
\newcommand{\datao}{\texttt{\href{http://doc.astro-wise.org/common.database.DataObject.html}{Data\-Object}}}
\newcommand{\dbobj}{\texttt{\href{http://doc.astro-wise.org/common.database.DBMain.html}{DBObject}}}

\journalname{Experimental Astronomy}
\begin{document}

\title{The \aw\ Optical Image Pipeline
}
\subtitle{Development and Implementation}

\titlerunning{\aw\ Optical Image Pipeline}        

\author{
        McFarland, J.P.       \and
        Verdoes-Kleijn, G.    \and
        Sikkema, G.           \and
        Helmich, E.M.         \and
        Boxhoorn, D.R.        \and
        Valentijn, E.A.
}

\authorrunning{McFarland, et al.} 

\institute{OmegaCEN, Kapteyn Astronomical Institute, Groningen University
           Postbus 800, 9700 AV, Groningen, The Netherlands\\
           \email{mcfarland@astro.rug.nl}
          }

\date{Received: date / Accepted: date}

\maketitle

\begin{abstract}

We have designed and implemented a novel way to process wide-field astronomical
data within a distributed environment of hardware resources and humanpower.
The system is characterized by integration of archiving, calibration, and
post-calibration analysis of data from raw, through intermediate, to final data
products.  It is a true integration thanks to complete linking of data lineage
from the final catalogs back to the raw data.  This paper describes the
pipeline processing of optical wide-field astronomical data from the
WFI\footnote{\url{http://www.eso.org/lasilla/instruments/wfi/}} and
OmegaCAM\footnote{\url{http://www.astro-wise.org/~omegacam/}} instruments using
the \aw\ information system (the \awenv\ or simply \awe).  This information
system is an environment of hardware resources and humanpower distributed over
Europe.  \awe\ is characterized by integration of archiving, data calibration,
post-calibration analysis, and archiving of raw, intermediate, and final data
products.  The true integration enables a complete data processing cycle from
the raw data up to the publication of science-ready catalogs.  The advantages
of this system for very large datasets are in the areas of: survey operations
management, quality control, calibration analyses, and massive processing.


\end{abstract}

\section{Introduction}

The rapid increase in the number of astronomical data sets and even faster
increase of overall data volume demands a new paradigm for the scientific
exploitation of optical and near-infrared imaging surveys.  Historical surveys
have been digitized (POSS and its southern counterpart) or are in the process
of being digitized\footnote{See,~e.g.,~\url{http://archive.stsci.edu/dss/},~\\
\url{http://tdc-www.harvard.edu/plates/},~\\
\url{http://www.lsw.uni-heidelberg.de/projects/scanproject/}}.  In recent
years surveys have been performed which cover hundreds or thousands of square
degrees up to the whole sky (SDSS, 2MASS, CFHTLS, etc.).  Many more are in
progress or coming up with increasing spatial resolution, depth, and survey
areas (OmegaCAM on VST, VIRCAM on VISTA, Pan-STARRS, LSST, etc.).  The data
rate of existing surveys is rapidly approaching \textit{terabytes} per night,
leading to survey volumes well into the \textit{petabyte} regime and the new
surveys will add many tens of petabytes to this\footnote{See,~e.g.,~\url{http://www.lsst.org/lsst/science/technology},~\\
\url{http://pan-starrs.ifa.hawaii.edu/public/design-features/data-handling.html}}.
Hundreds of terabytes of data will start entering the system when ESO's
OmegaCAM camera starts operations in Chile in late-2011.  Several large surveys
plan to use the \aw\ information system to manage their data: the 1500 deg$^2$
\kids\ Survey\footnote{\url{http://www.astro-wise.org/projects/KIDS/}}, the
Vesuvio Survey\footnote{\url{http://www.astro-wise.org/projects/VESUVIO/}} of
nearby superclusters, the
OmegaWhite\footnote{\url{http://www.astro-wise.org/projects/OMEGAWHITE/}} white
dwarf binary survey and the
OmegaTrans\footnote{\url{http://www.astro-wise.org/projects/OMEGATRANS/}}
search for transiting variables.

Quality control is typically one of the largest challenges in the chain from
raw data of the ``sensor networks'' to scientific papers.  It requires an
environment in which all non-manual qualification is automated and the
scientist can graphically inspect where needed by easily going back and forth
through the data (the pixels) and metadata (everything else) of the whole
processing chain for large numbers of data products.  The full quality control
mechanisms are treated in complete detail in the \aw\ Quality Control paper
\citep{mcfarland}.

The really novel aspect of this new paradigm is the long-term preservation of
the raw data and the ability of re-calibrating it to the requirements of new
science cases.  The data of the majority of these surveys is fully public: any
astronomer is entitled to a copy of the
data\footnote{See,~e.g.,~\url{http://www.eso.org/sci/observing/policies/PublicSurveys.html}}.
Therefore the same survey data is used for not only science cases within the
original plan, but many new science cases the original designers of the survey
were not planning to do themselves or did not foresee.  To be able to do this
successfully requires that everyone is provided access to detailed information
on the existing calibration procedures and resulting quality of the data at
every stage of the processing, that is, have access to the data and the
metadata, including process configuration at every step in the chain from raw
data to final data products.

In this paper we describe the reduction of data in the \aw\ information
system, generally referred to as the \awenv\ (hereafter \awe).  The processing
of data from both the WFI and OmegaCAM instruments has been used to qualify the
pipeline, the results of which have been or will be included in separate
publications, for example \citet{verdoes,omegacam}.  The remainder of this
section briefly describes some key concepts of \awe\ covered in detail
elsewhere: previously in \citet{adass} and more currently in \citet{begeman}.
Sections~\ref{sec:calib}~and~\ref{sec:image} describe how an instrument is
calibrated and how science data is processed.  Finally, Sect.~\ref{sec:summa}
presents the summary.

\subsection{Context}\label{sec:conte}

Context is the primary tool of project managers in \awe.  Each \textit{process
target} (i.e., the result of some processing step, see Sect.~\ref{sec:objec})
in \awe\ is created at a specific privilege level.  Privilege levels are
analogous to the permission levels of a Unix/Linux file system (e.g., privilege
levels 1, 2, 3 map loosely to permission levels user, group, other).  To allow
access to their desired set of objects, users can set their privilege level and
their project.

This concept of \textit{context} is completely about visibility of the objects
in \awe\ and nothing else.  Proprietary data is protected from access by all
but authorized users and undesirable data can be hidden for any purpose (e.g.,
to use project-specific calibrations instead of general ones).  All processing
is done within this framework, allowing complete control over what is processed
and how, and how it is \textit{published} between project groups and to the
world.

Visibility for processing targets is not only governed by the privilege level,
but also by validity.  Three properties dictate validity:
\begin{enumerate}
\item
\texttt{is\_valid} -- manual validity flag
\item
\texttt{quality\_flags} -- automatic validity flag
\item
\textit{timestamps} -- validity ranges in time (for calibrations only)
\end{enumerate}
Determining what needs to be processed and how is indicated by setting any or
all of the above flags.  For instance, obviously poor quality data can be
flagged by setting its \texttt{is\_valid} flag to 0, preventing it from ever
being processed automatically.   The calibrations used are determined by their
\textit{timestamps} (Which calibrations are valid for the given data?) and the
quality of processed data by the automatic setting of its
\texttt{quality\_flag} (Is the given data good enough?).  Good quality data can
then be flagged for promotion (\texttt{is\_valid} $>1$) and eventually promoted
in privilege by its creator (published from level 1 to 2) so it can be seen by
the project manager who will decide if it is worthy to be promoted once again
(published from level 2 to 3 or higher) to be seen by the greater community.

\subsection{Provenance: full dependency linking}\label{sec:prove}

\awe\ uses its federated database to link all data products to their
progenitors (dependencies), creating a full data lineage of the entire
processing chain.  This allows creation of complete data provenance for any
data item in the system at any time.

\subsubsection{Full data lineage}

Raw data is linked to the final data product via database links within the
\textit{data object}, allowing all information about any piece of data to be
accessed instantly.  See \citet{mwebaze} for a detailed description of the
\awe's data lineage implementation.  This data linking uses the power of
Object-Oriented Programming to create this framework in a natural and
transparent way.

\begin{figure}
\begin{center}
\includegraphics[angle=0,width=118mm]{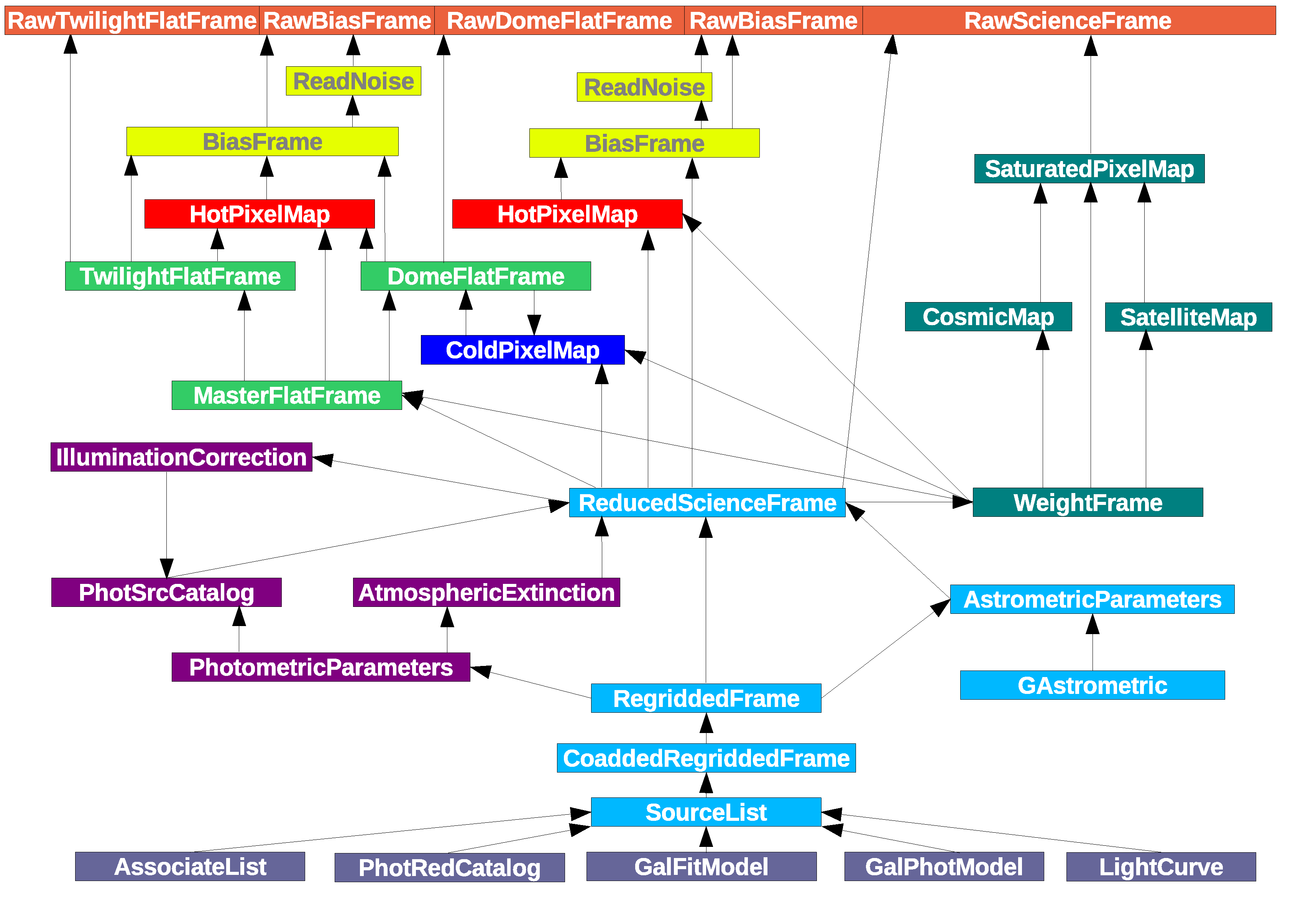}
\caption{
A \textit{target diagram}: slightly simplified object model that is a view of
the dependencies of ``targets'' to raw observational data.  The arrows indicate
the backward chaining to the raw data, not the progression through any
processing pipeline.  The colors provide a visual grouping of similar types of
data products.
}\label{fig:targe}
\end{center}
\end{figure}

\begin{figure}
\begin{center}
\includegraphics[angle=0,width=94mm]{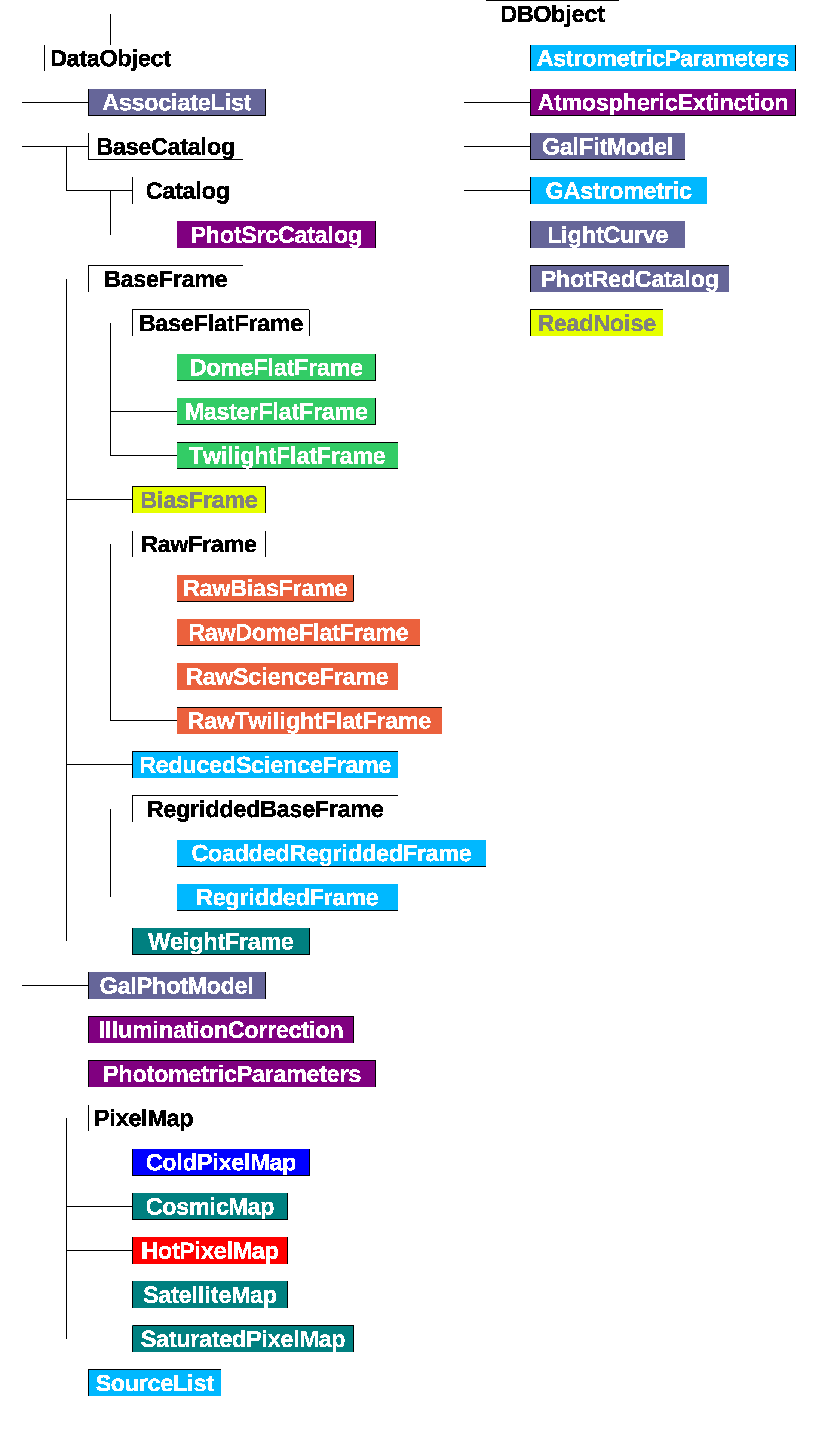}
\caption{
A \textit{\aw\ hierarchical object model}.  A simplified object model of the
target \textit{classes} shown in Fig.~\ref{fig:targe} illustrating their
inheritance relationship to each other.  The classes without color do not
appear in the previous figure, but are nonetheless part of the hierarchy and
are shown for clarity.  Every target inherits from \dbobj\ (a database object),
but only those with associated bulk data (typically a file stored on a
dataserver) inherit from \datao.
}\label{fig:class}
\end{center}
\end{figure}

\subsubsection{Object-oriented data model}\label{sec:objec}

\awe\ uses the advantages of Object-Oriented Programming (OOP) to process data
in the simplest and most powerful ways.  In essence, it turns the
aforementioned \textit{data objects} into OOP \textit{objects}, called
\textit{process targets} (or {\proce}s), that are instances of classes with
attributes and methods that can be inherited (see
Fig.~\ref{fig:targe}~and~\ref{fig:class} for an overview of an \aw\ object
model).  Each of these \proce\ instances knows of all of its local and linked
metadata, and knows how to process itself.  Each persistent attribute of an
object is linked to metadata or to another object that itself contains links to
its own metadata.

The code for \awe\ is written in \py, a programming language highly suitable
for OOP. Consequently, \py\ \textit{classes} are associated with the
various conventional calibration images, data images, and other derived data
products.  For example, in \awe, bias exposures become \textit{instances}
of the \rawbi\ class, and twilight (sky) flats become instances of the \rawtw\
class.  These instances of classes are the ``objects'' of OOP.

For the remainder of this document, the class names of objects, their
properties, and methods will be in \texttt{teletype} font for more clear
identification.

\subsection{Target-based processing}\label{sec:targe}

The most unique aspect of \awe\ is its ability to process data based on
the final desired result to an arbitrary depth.  In other words, the data is
\textit{pulled} from the system by the user.  The desired result is the
\textit{target} to be processed, and the framework used is called
\textit{target processing}.  Target processing uses methods similar to those
found in the Unix/Linux \texttt{make} utility.  When a target is requested, its
dependencies are checked to see if they are \textit{up-to-date}.  If there is a
newer dependency or if the requested target does not exist, the target is
(re)made.  This process is recursive and is an example of \textit{backward
chaining}.

\subsubsection{Backward chaining}

At the base of \awe\ target processing is the concept of \textit{backward
chaining}.  Contrary to the typical case of forward chaining (e.g.,
\texttt{objectN} is processed into \texttt{objectN+1} is processed into
\texttt{objectN+2}, etc.).  \awe\ database links allow the dependency chain to
be examined from the intended target (even if it does not yet exist) all the
way back to the raw data.  The above scenario would then look like:
if \texttt{targetM} is up-to-date, check
if \texttt{targetM-1} is up-to-date; if \texttt{targetM-1} is up-to-date, check
if \texttt{targetM-2} is up-to-date; etc., processing as necessary until
%
%
\texttt{targetM} (and all targets it depends on) exists and is
up-to-date\footnote{Note that the counting of targets is reversed in the
backward chaining example, as this is the direction in which the up-to-date
check is run.}.  This is the \awe\ implementation of backward chaining that is
used in target processing (see Fig.~\ref{fig:targe} for an example with
astronomical data).

\subsubsection{Processing parameters}

As mentioned earlier, conventional astronomical calibration images/products as
well as science products are collectively referred to as \textit{process
targets} and inherit from the \proce\ class.  Each \proce\ has an associated
\textit{processing parameters} object, an instance of a class named after the
respective process target class (e.g.,
\texttt{SomeTarget.SomeTargetParameters}) which stores configurable parameters
that guide the processing or reprocessing of that target.  Those {\proce}s that
use external programs in their derivation may have additional objects
associated with them which contain the configuration of the external program
that was used.

These processing parameters are stored in an object linked to the \proce\ for
comparison by the system and to allow the all persons involved in survey
operations to discover which settings resulted in the best data reduction.

\begin{figure}
\begin{center}
\includegraphics[angle=0,width=118mm]{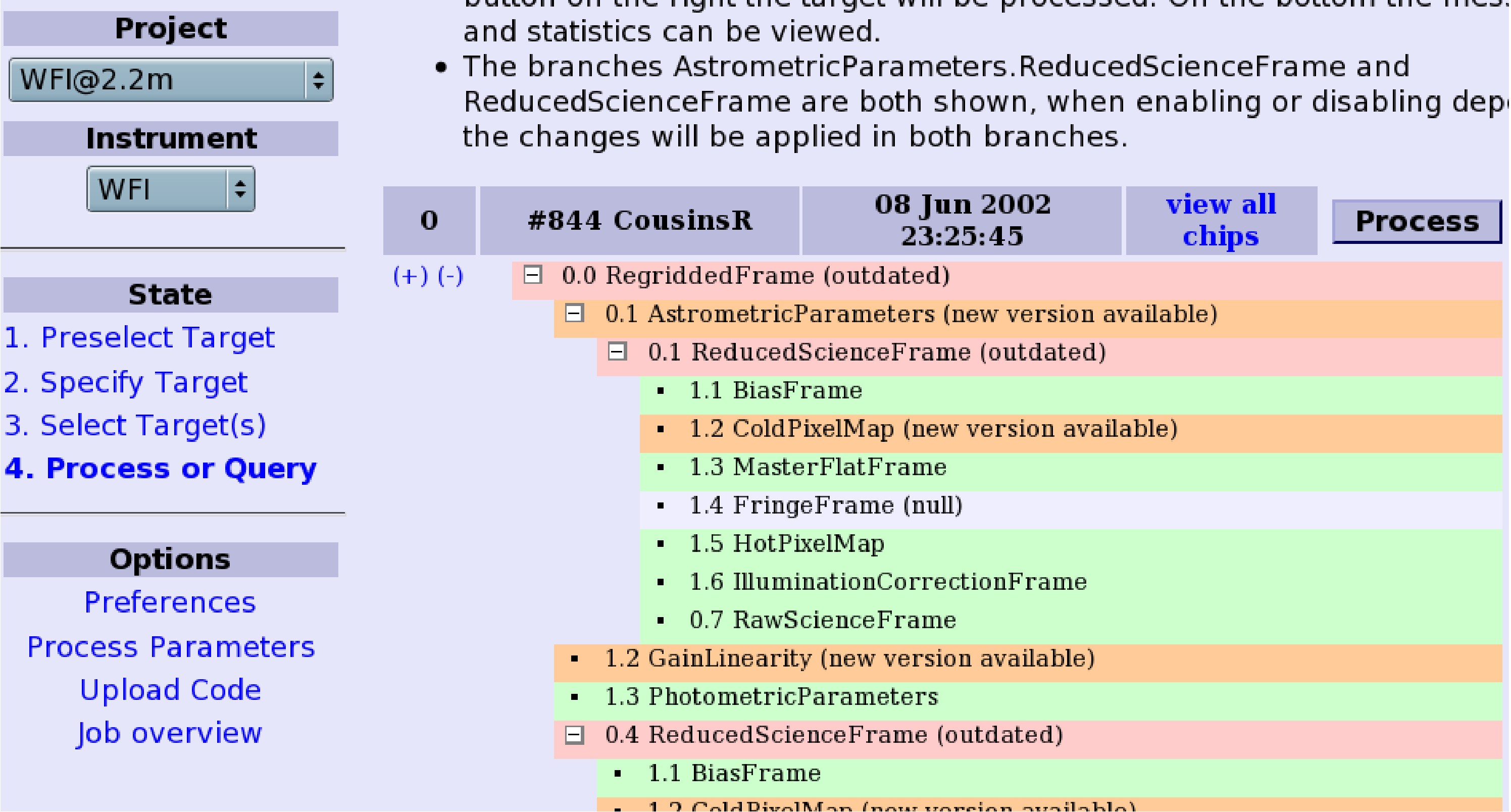}
\caption{
A screen-capture of part of the web-based \textit{target processing} interface.
On the left are high-level processing settings (e.g., project, processing step,
options).  On the right is the result of the query for a particular target.
Green rows show dependencies that are ready and will not be processed, red and
orange rows show dependencies that are either outdated (need to be rebuilt) or
already have a new version available.  This section is a glimpse at the
information used to dynamically construct the workflow that will create the
eventual processing pipeline.  Only those \textit{targets} in the red rows will
actually be processed.
}\label{fig:tproc}
\end{center}
\end{figure}

\subsection{On-demand reprocessing}\label{sec:ondem}

\awe\ combines all of the above concepts into a coherent archiving and
processing system.  All the information about a particular instrument and its
calibration and processing history is stored in the federated database within
the object-oriented data model with full linking of the data lineage.  The
values of the process parameters of all objects in the dependency chain and all
the results of the integrated (and manual) quality controls of the
\textit{target} of interest (regardless of visibility or existence) are used to
determine if that \textit{target} can or should be (re)built and how.  This
data pulling is the heart of \awe\ and is called \textit{target processing}
(see Fig.~\ref{fig:tproc} and \url{http://process.astro-wise.org/}).

\subsubsection{Raw data sacred}

As mentioned earlier, \awe\ does not provide as the ultimate end of the
processing chain a static data release.  The system allows for survey data to
be reprocessed for any reason and for any purpose.  If a newer, better
calibration is made, or if a different purpose requires a different processing
technique, the data can be easily reprocessed.  This is only possible when the
raw survey data is retained in its original form.  In \awe\, raw data is
always preserved.

\subsubsection{On-the-fly (re)processing}

Target processing does not use static information to determine what gets
processed how.  As seen in all the previous sections, all the survey data, its
dependency linkages and processing parameters are all reviewed to allow any
target to be (re)processed on-demand as needed.  All these dependencies create
a built-in workflow, automatically processing only those targets that need it.
This on-the-fly (re)processing is the hallmark of the \awe\ information system.


\section{Calibration Pipeline: correcting the pixels}\label{sec:calib}

\begin{figure}
\begin{center}
\includegraphics[angle=0,width=118mm]{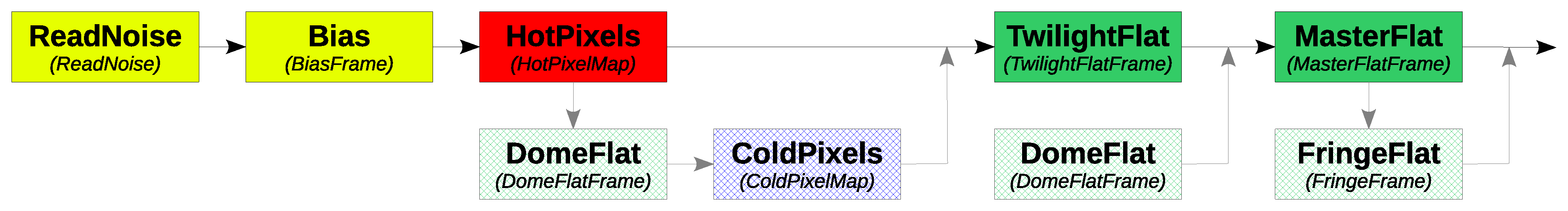}
\caption{
Schematic flow of the pixel calibrations pipeline following the coloring in Fig.~\ref{fig:targe}.  The recipes, also called \texttt{Tasks}, used to produce
various {\proce}s are indicated in each box (with their data product in
parentheses) and described in the various sections.  The arrows connecting them
indicate the direction of processing.  Note that the sections with the hatched
boxes are optional branches in this pipeline, and the arrow at the end leads to
the beginning of the photometric pipeline schematic in Fig.~\ref{fig:photo}.
Also note, in order to simplify this diagram, the \gainl, \darkc\ and \night\
objects have been omitted.
}\label{fig:calib}
\end{center}
\end{figure}

\begin{figure}
\begin{center}
\includegraphics[angle=0,width=118mm]{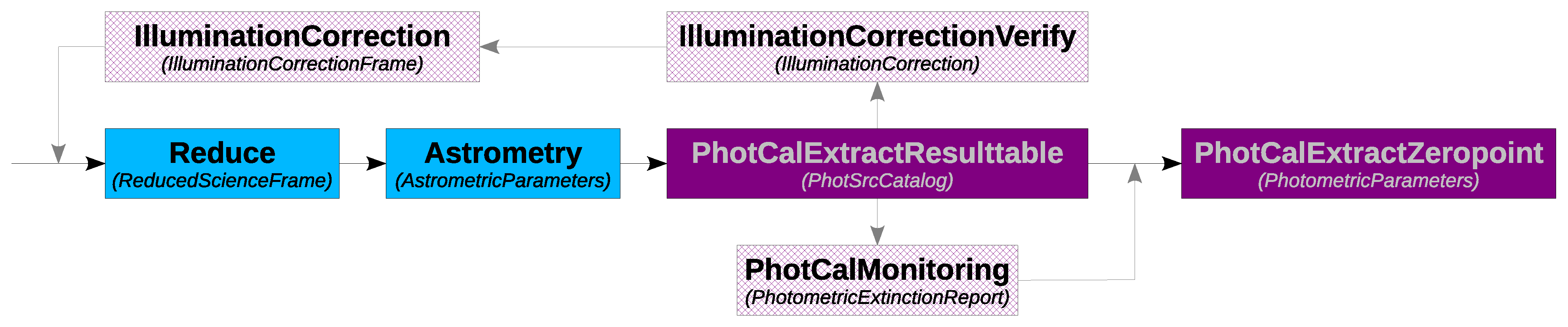}
\caption{
Schematic flow of the photometric pipeline following the coloring in Fig.~\ref{fig:targe}.  The recipes, also called \texttt{Tasks}, used to produce
various {\proce}s are indicated in each box (with their data product in
parentheses) and described in the various sections.  The arrows connecting them
indicate the direction of processing.  Note that the sections with the hatched
boxes are optional branches in this pipeline, and the input follows from the
pixel calibrations pipeline shown in Fig.~\ref{fig:calib}.
}\label{fig:photo}
\end{center}
\end{figure}

The philosophy of \awe\ is to share improved insight in calibrations.  In \awe,
calibration scientists can, over time, have many versions of calibration results at their
disposal.  From this they determine (subtle) long term trends in instrument,
telescope and atmospheric behaviour and can collaborate to improve the
calibration procedures for that instrument in \awe\ accordingly.  The
\textit{complete observational system} (generally termed ``the instrument'' for
simplicity) eventually becomes calibrated over its full operational period as
opposed to a series of individual nights calibrated from data in a limited time
window.  Fig.~\ref{fig:calib} shows the schematic view of the pixel
calibrations pipeline.  This gives an overview of the flow of the pixel
calibrations to be described in the coming sections.  It is continued in the
photometric pipeline schematic in Fig.~\ref{fig:photo}.

\begin{figure}
\begin{center}
\includegraphics[angle=0,width=118mm]{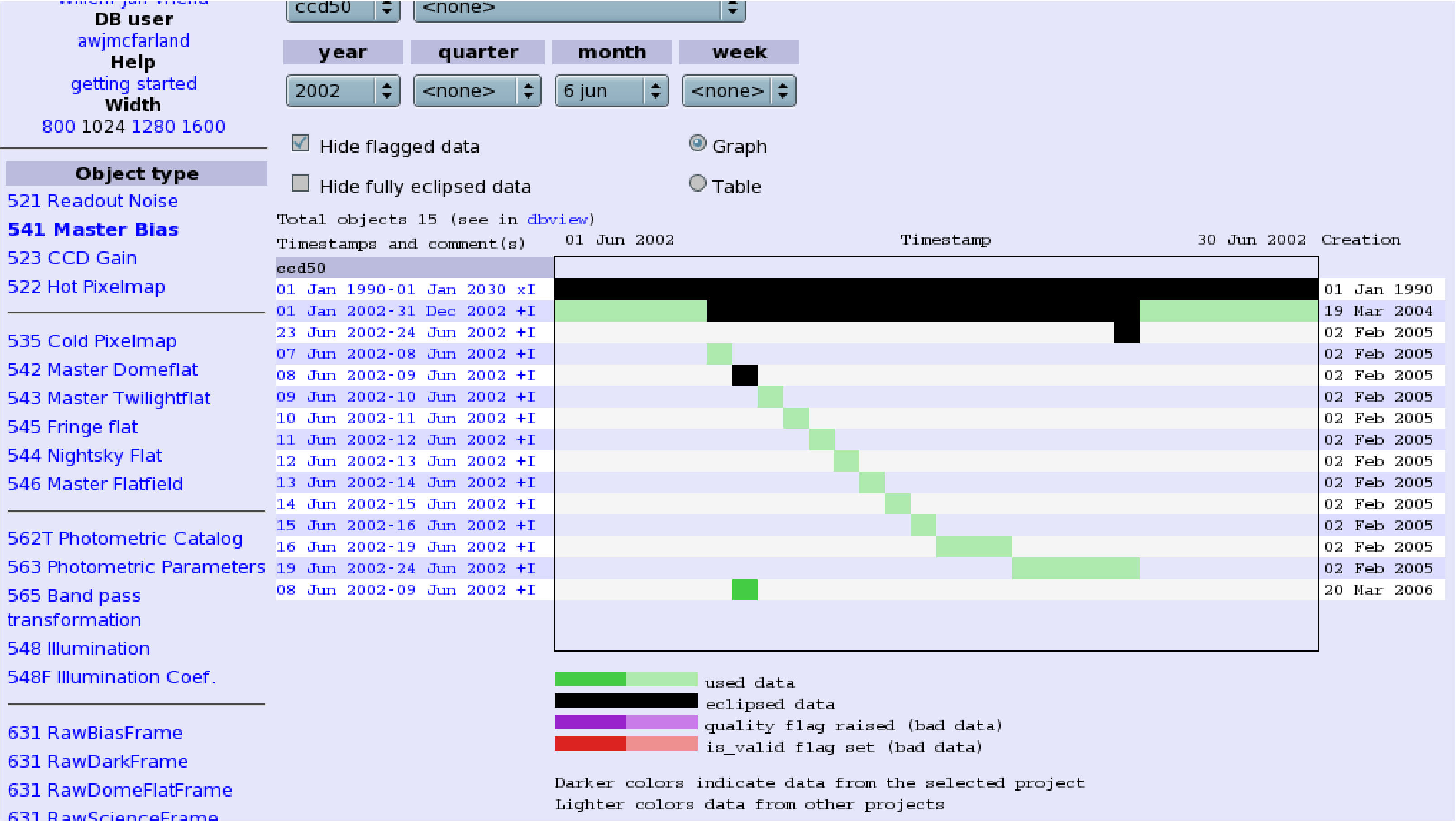}
\caption{
A screen-capture of CalTS, the web-based Calibration TimeStamp service.  The
purpose of this service is to give a graphical representation of the temporal
validity ranges of calibration objects in \awe.  On the left can be selected
the \proce\ of interest, at the top are some of the query criteria, and below
this, the graphical validity of the \proce.  Colored bars indicate the most
recent valid objects (objects flagged invalid are hidden), while black bars
indicate where objects are ``eclipsed'' by newer calibrations.  It is always
assumed that the newest valid \proce\ is the best and this will be the one used
during processing.  The timestamps and validity can be modified by an interface
raised by clicking on the date range for a given object.  \url{http://calts.astro-wise.org/}
}\label{fig:calts}
\end{center}
\end{figure}

In the \awe, calibration objects have a set validity range in time or per frame
object that depends upon the calibration object (the defaults are specified per
calibration object in Table~\ref{tab:valid} below).  The default validity time
range (\texttt{timestamp\_start} to \texttt{timestamp\_end}) can be altered on the
command-line using context methods (see Sect.~\ref{sec:conte}), or via the CalTS
web-service (see Fig.~\ref{fig:calts}).

\begin{table}
\begin{center}
\begin{tabular}{|l|l|}
\hline
\textbf{ProcessTarget}          & \textbf{Default validity}             \\
\hline
GainLinearity                   & 1 day                                 \\
ReadNoise                       & 1 day                                 \\
BiasFrame                       & 1 day \\
DarkCurrent                     & 1 day                                 \\
HotPixelMap                     & same as source BiasFrame              \\
ColdPixelMap                    & same as source FlatFrame              \\
DomeFlatFrame                   & 7 days                                \\
TwilightFlatFrame               & 7 days                                \\
NightSkyFlatFrame               & 1 day                                 \\
MasterFlatFrame                 & 7 days \\
FringeFrame                     & 1 day                                 \\
AstrometricParameters           & points to one frame only              \\
AtmosphericExtinctionCurve      & 1 day                                 \\
PhotometricReport               & 1 day                                 \\
PhotometricParameters           & 1 day                                 \\
IlluminationCorrection          & 1 day                                 \\
IlluminationCorrectionFrame     & same as source                        \\
                                & IlluminationCorrection                \\
\hline
\end{tabular}
\end{center}
\caption{
Default validities of calibration {\proce}s.  All time spans are centered on
local midnight of the day the source observations were taken unless otherwise
indicated.
}\label{tab:valid}
\end{table}

\begin{table}
\begin{center}
\begin{tabular}{|l|l|r|r|}
\hline
\textbf{Class} & \textbf{process\_param}     & \textbf{value} & \textbf{units}\\
\hline
\readn & rejection\_threshold                &           5.0  &           \\
       & maximum\_iterations                 &           5    &           \\
\hline
\gainl & overscan\_correction                &           6    &           \\
       & rejection\_threshold                &           5.0  &           \\
       & maximum\_iterations                 &           5    &           \\
\hline
\biasf & overscan\_correction                &           6    &           \\
       & sigma\_clip                         &           3.0  &           \\
\hline
\hotpi & rejection\_threshold                &           5.0  &           \\
       & maximum\_iterations                 &           5    &           \\
\hline
\coldp & threshold\_low                      &           0.94 &           \\
       & threshold\_high                     &           1.06 &           \\
\hline
\domef & overscan\_correction                &           6    &           \\
       & sigma\_clip                         &           3.0  &           \\
\hline
\twili & overscan\_correction                &           6    &           \\
       & sigma\_clip                         &           3.0  &           \\
\hline
\maste & dig\_filter\_size                   &           9.0  &           \\
       & mirror\_xpix                        &          75    & pixel     \\
       & mirror\_ypix                        &         150    & pixel     \\
       & median\_filter\_size                &          36    & pixel     \\
       & combine\_type                       &           1    &           \\
\hline
\photo & sigclip\_level                      &           1.5  &           \\
       & min\_nmbr\_of\_stars                &           3    &           \\
\hline
\end{tabular}
\end{center}
\caption{
Processing parameters and their generic default values.  These values are
representative of the typical value for any instrument.  Some instruments may
have values that different from these based on experience with that instrument.
See the document page linked from the class name or appropriate links on
\url{http://doc.astro-wise.org/astro.main.html} for more details.
}\label{tab:calib}
\end{table}

Be sure to note that, with the exception of parts of the astrometric
calibration derivation and most of the photometric calibration derivation, all
calibration objects are normally processed in a parallel environment, one
detector chip per CPU node.

Many \proce's have configurable \textit{processing parameters} to control how
they are processed.  Table~\ref{tab:calib} gives an overview of these
\texttt{process\_params} for the calibration pipeline.  In addition to the
\texttt{process\_params} associated directly with the \proce, there exist
object representations of configuration files for external programs wrapped in
\py\ (e.g., SExtractor, SWarp, etc.).

\subsection{\readn}

The read-out noise is the noise introduced in the data by the read-out process
of detector chips.  It is measured from pairs of bias exposures.  The RMS
scatter of the differences between two bias exposures is computed. The read
noise in ADU is determined via division of this value by $\sqrt{2}$.  The
read noise value is stored in the database using the \readn\ class.

\subsection{\gainl}

The gain is the conversion factor between the signal in ADU's supplied by the
readout electronics and the detected number of photons (in units e$^{-}$/ADU).
For OmegaCAM, a procedure (template) to determine the gain (and the linearity
of the detector chips) is defined that involves taking two series of 10 dome
flatfield exposures with a wide range of exposure times, and deriving the RMS
of the differences of two exposures taken with similar exposure (integration
time).  The regression of the square of these values with the median level
yields the conversion factor in e$^{-}$/ADU (assuming noise dominated by photon
shot noise).  A linear fit of exptimes vs. median\_sum gives a measure of the
linearity.  For most instruments default gain values have been determined or
taken from the literature and are in the system, so it is usually not necessary
to make new values for them.  If this is desired, a specialized dataset similar
to that described must be used.  The class used to store the gain in the
database is the \gainl.

\subsection{\biasf}

The signal in raw scientific frames contains a component that is due to a bias
current introduced by the AD converter on a FIERA\footnote{Acronym for Fast
Imager Electronic Readout Assembly CCD controller,
\url{http://www.eso.org/projects/odt/Fiera/}} or other detector controller.
This component shows up as an offset to the signal.  In most CCD detectors, the
bias-offset has the following characteristics: i) the bias level grows to its
asymptotic level in the first few hundred lines, and ii) the bias level depends
on the total signal in a given line.  Therefore, an initial bias
correction--the \textit{overscan correction}, is applied when the overscan
region exists (cheaper CCDs and IR detectors tend not to have these regions).
The method used is one of a set of methods ranging from no correction, to
subtraction of a constant value derived from one of the prescan or overscan
regions, to subtracting an average value per column or row, smoothed or not, to
hybrid corrections for complex geometries.  Each of these methods is given an
index which is stored in the database, constituting the only really ``free''
parameter in the system.

In addition, the bias offset exhibits a residual pattern, which is measured by
the master bias frame, an instance of the \biasf\ class.  To construct the
master bias, a series of $N$ (usually 5-10) zero-second bias exposures is
overscan-corrected and averaged, rejecting $5\sigma$ outliers ($\sigma=$
readout noise from a \readn\ object), due to particle hits during read-out.
The resulting master bias frames will be used for the correction of all frames.

As the read-out noise dominates the RMS scatter in the bias frames, while
the shot noise of the sky background dominates the RMS scatter on the sky
images, which is nominally much larger than the readout noise, it is
sufficient to characterize the bias value at individual pixels with an
accuracy of (readout noise~$/\sqrt{N}$).

\subsection{\darkc}

In \awe, no formal dark frame subtraction is performed.  Current, liquid
nitrogen cooled instruments tend to have little or no appreciable
two-dimensional dark current structure, any of which will normally be removed
with the sky background.  As \awe\ was created explicitly for such an
instrument, dark frame correction was not included.  There is, however, some
treatment of this effect through the \darkc\ class.  The purpose of this class
is to determine the total dark current and the particle event rate of a
detector chip.  This is not used for calibration, but for the detector chain
health.

The dark current, excess signal due to heat in a detector chip, is measured by
taking 3 identically timed exposures (typically one hour) with the camera
shutter closed.  The resulting frames are trimmed, overscan- and
bias-corrected, then a median is taken along the Z-axis of the exposure stack.
After iterative outlier rejection, the average value of all the pixels is the
dark current in units of ADU/pixel/hour.

The same trimmed, overscan- and bias-corrected frames are used to determine the
particle even rate.  The source extraction software
SExtractor\footnote{\url{http://astromatic.iap.fr/software/sextractor/}} is
used on each image in turn to detect the number of cosmic ray particle events.
A \hotpi\ can optionally be used to mask detected hot pixels.  The particle event
rate is determined in units of particles/cm$^2$/hour.

\subsection{\hotpi\ and \coldp}

Hot pixels are pixels which have high count rates despite not being
illuminated.  In \awe, these pixels are detected from bias images (which
have an exposure time of 0 seconds).  More precisely: greater than $5 \sigma$
outliers in bias are defined as hot pixels.  Cold pixels are broken pixels
which have low or zero counts even when illuminated.  These pixels are
determined from dome flat-field exposures because those have the most
uniform and consistently high counts required.  Twilight flat-fields
can be used if no dome flat-fields are available.  In \awe, all pixels that
deviate substantially, i.e., more than 4\% of its surroundings, from the
other pixels in the flat-field are considered cold even though brighter
pixels are also detected.  All deviant pixels are flagged in weight maps, a
\textit{mask image}, where good pixels have a value of 1 and bad pixels a
value of 0.

The procedure to create a \hotpi\ starts with calculating a background map of
the master bias frame and subtracting it.  This is done to avoid detecting
induced charge structures and other continuous structures as hot pixels.
Outliers in the background-subtracted master bias frame are bad/hot pixels.  A
\hotpi\ is created using the threshold determined from iterative statistics
estimates.  The number of hot pixels is noted as a quality control value.

The procedure to create a \coldp\ starts with smoothing the flat-field image.
The smoothed flat is used to normalize, or ``flatten'' the flat to eliminate
large deviations from flatness that could erroneously cause entire regions to
be marked as ``cold''.  In this flat-field image, pixels that are outside a
given range ($\pm 4\%$) are taken to be cold pixels.  Note that this
invalidates any pixel whose gain differs significantly from its immediate
neighbors.  In particular, this also identifies pixels that are bright relative
to their neighbors as ``cold''.  Note, that pixels above the threshold are
formally not \textit{cold}, but are flagged anyway.  In the end, {\hotpi}s and
{\coldp}s are combined into weights of the detrended science images.  A \coldp\
is created using the thresholds given above.  The number of cold pixels is
noted as a quality control value.

We use SExtractor to produce the smoothed images.  SExtractor uses a robust
algorithm to estimate the background on a grid and interpolate between these
grid points. By measuring this background for the bias and flat-field we
essentially have a fast smoothing algorithm with a large kernel, that is
relatively insensitive to bad pixels.

\subsection{Flat fielding}

A flat-field is the response of the telescope-camera system to a source of
uniform radiation.  In \awe, there are different ways to construct a flat
field.  Dome flat-fields are created by pointing the telescope at a screen on
the inside of the dome which is illuminated by lamps.  Dome flat fields have
the advantage (over twilight flat fields) that it is easy to repeatedly obtain
a high signal to noise level.  Disadvantages are that the direction in which
light enters the telescope may be different than during night time
observations, that the color of the dome lamp differs from the color of the
night sky and that it is very difficult to illuminate a screen in such a way
that it is a source of uniform radiation.  A dome flat field is useful for
tracing small scale structure variations.
A disadvantage for twilight flats is that they can already contain objects like
stars during exposures, which should be corrected for by dithering the twilight
flats.  Twilight flat fields thus are better in tracing large scale structure
variations.  These considerations result in the desire to combine dome flats
and twilight flats by spatially filtering the two types of flat fields.

\subsubsection{\domef}

A \domef\ is obtained through an average with sigma rejection
procedure on a stack of raw dome flats, intended to reduce photon shot
noise and remove cosmic rays.

The procedure to make a \domef\ starts with 5-10 overscan corrected, trimmed
and debiased raw dome flats.  These are normalized to the median, taking into
account hot and cold pixels, and averaged rejecting $5\sigma$ outliers: the
median in Z-axis of the stack is used to determine the $\sigma$ levels.  The
computed mean in the Z-axis of the stack is the final \domef\ image.  Lastly,
sub-window image statistics are determined for quality control purposes.

\subsubsection{\twili}

A \twili\ is obtained through an average with sigma rejection
procedure on a stack of raw twilight flats, intended to remove any
contamination (including stars) present on individual raw twilight flats
and reduce photon shot noise.

The procedure to make a \twili\ starts with 5-10 overscan-corrected, trimmed
and debiased raw dome flats.  These are normalized to the median, taking into
account hot and cold pixels, and averaged rejecting $5\sigma$ outliers: the
median in Z-axis of the stack is used to determine the $\sigma$ levels.  The
computed mean in the Z-axis of the stack is the final \twili\ image.  Lastly,
sub-window image statistics are determined for quality control purposes.

\subsubsection{\night}

Raw science images have a non-flat background, attributed to flat-field
effects.  Information about how to flat-field science images therefore is
present in the science images themselves.  The flat-field that most closely
reproduces the actual gain variations of the these images can be obtained by
averaging a large number of \textit{flat-fielded} science and standard
observations, taking care of properly masking the contaminating objects.  Such
a night-sky flat could, in principle, improve on the quality of the twilight
flat and may also be suitable for fringe removal.

The procedure to create a \night\ starts with a minimum of 5 non-cospatial
science images within a given night in a given band to achieve optimal results.
Images are overscan-corrected, trimmed, debiased, flat-fielded and normalized,
then stacked and a median along the Z-axis is calculated.  This median is
intended to remove any exposure-specific effects (objects, cosmic rays,
satellite tracks, etc.).  The median image is then normalized to the mean
taking into account hot and cold pixels.

\subsubsection{\maste}\label{sec:maste}

In \awe, a \maste\ constructed from a \domef\ (used to measure the small-scale
pixel-to-pixel variation) and a \twili\ (used to measure the large-scale
variation).  These spatial frequencies are separated using a Fourier technique.
{\night}s are created from raw science or standard data that has been
flat-fielded with this master flat-field and can be used to improve the quality
of it.  This (improved) master flat-field is then used to flat-field the
science and standard images in the image pipeline.

In practice, not all three flat-field types are available.  As a result, \awe\
offers three different combination methods:
\begin{enumerate}
\item
the \maste\ is constructed by extracting high spatial frequency components from
the \domef\ and low spatial frequencies from the \twili, multiplied to give the
master flat
\item
the \maste\ is a direct copy of the \domef
\item
the \maste\ is a direct copy of the \twili
\end{enumerate}
In all cases a \night\ can be provided which is multiplied with this master
flat-field as an improvement on it as mentioned above.

In certain situations, it may be advantageous to split the \domef\ and \twili\
contributions out of the process.  The machinery of \awe\ allows this to be
accomplished in a straight-forward manner.  The advantages to this would be in
isolating either large-scale (low spatial frequencies) or small-scale (high
spatial frequencies), pixel-to-pixel variations of the \twili\ or \domef,
respectively.  This concept will be explored further in Sect.~\ref{sec:illum}.

To give a more detailed description, low spatial frequencies are extracted from
the master dome and master twilight flats by the process indicated below. The
high spatial frequencies of the dome flat are obtained by dividing the dome
flat by its low spatial frequency components. The low spatial frequencies of
the twilight flat are then multiplied by the high spatial frequencies of the
dome flat.

Low spatial frequencies are extracted as follows:
\begin{itemize}
\item
all bad pixels in input images are replaced by the median value of the pixels
in a box around the bad pixel
\item
to reduce problems with Fourier filtering near image edges the size of the
image is increased by mirroring the edges and corners
\item
a two-dimensional array is created containing the equivalent of a circular
Gaussian convolution function in Fourier space (taking into account the
quadrant shift introduced by the Fourier transform)
\item
the Fourier transform of the image is multiplied by the Gaussian filter
\item
the image is transformed back, and the mirrored regions removed
\item
the resulting image is normalized, excluding bad pixel values
\end{itemize}

\subsubsection{\fring}

Fringing requires a different approach to background subtraction.  Fringing in
a solid state detector chip is due to interference of incident photons with
photons reflected in the detector chip substrate.  The photons causing the
strongest fringes are those of several skylines, mostly apparent at the long
wavelengths, that can vary with filter.  Normally, after flatfielding, the
background can be expected to be flat over the entire image, and a median of
the image, excluding $5\sigma$ outliers, would in principle be sufficient to
subtract the background.

In images that suffer from fringing we have to deal with a background that is
variable on small ($\ll1'$) scales within the image, and can not be
distinguished from sources.  The image itself can, therefore, not be used to
determine the background.  However,
the information of several images can be combined to determine a background.
This average should include enough observations to properly exclude
contamination from sources.

A suitable strategy to construct a fringed background image, usable for
subtraction, thereby removing the fringe pattern, remains to be determined. If
the fringe pattern is stable over the night, a decomposition of the night-sky
flat in an additive and multiplicative term is feasible. The assumption that
the high-frequency spatial component in the night-sky flat are fringes, while
the lowest frequency components represent gain variations has been used with
reasonable success.

The procedure to create a \fring\ starts with a minimum of 3 non-cospatial
science images of reasonably long exposure time (e.g., greater than 30 seconds) within a
given night in a given band to achieve optimal results.  Images are
overscan-corrected, trimmed, debiased, flat-fielded and normalized, then
stacked and a median along the Z-axis is calculated.  This median is intended
to remove any non-systematic effects (objects, cosmic rays, satellite tracks,
etc.).  The median image is then normalized to the mean taking into account hot
and cold pixels.  The value of 1.0 is subtracted from the normalized fringe map
to obtain an average value of zero.  Bad pixels are assigned a value of zero by
multiplying by the combined hot and cold pixel maps.

During a night the brightness of the emission lines will change, especially
near evening and morning twilight. The result of this is that the amplitude of
the observed fringes will change. Therefore, fringe maps should be scaled to
fit the amplitude of the fringes in each science frame. This is calculated from
the standard deviation in a science image, which is derived from all non-bad
pixels that have values within a given threshold from the median background
level. It is assumed that this standard deviation depends on the amplitude of
the fringes.

\subsection{\astro}\label{sec:astro}

Astrometric calibration is a vital, integral part of any astronomical data
reduction and analysis system.  \awe\ performs two kinds of astrometric
calibration of pixel data.  Their results are termed \textit{local astrometry}
and \textit{global astrometry}.  The goal of the global astrometry is to improve on
the local astrometry.  Unlike all the previous calibrations, the resulting \astro\
objects are each linked to a single processed science observation (a single
detector chip of one exposure), as it is that observation that provides the
source positions to be calibrated via the astrometric solution.

The local astrometric solution (see Sect.~\ref{sec:local}) is derived on the
basis of a single detector chip's information.  It is obtained by minimizing
the differences between the RA and DEC positions of sources in a single
detector chip and their positions listed in a catalog of astrometric standards.
The global astrometric solution (see Sect.~\ref{sec:globa}) can be obtained
if one has dithered (nearly cospatial and cotemporal) observations and local
astrometric solutions for each detector chip.  It then additionally minimizes
the positional differences of sources appearing on more than one detector chip.
This results in a higher accuracy of the astrometric calibration.  The use of
global astrometry improves the image quality of a coaddition of dithered
observations compared to local astrometry.

In \awe, astrometric solutions are solved by running \texttt{LDAC} (Leiden Data
Analysis
Center\footnote{\href{ftp://ftp.strw.leidenuniv.nl/pub/ldac/software/pipeline.pdf}{ftp://ftp.strw.leidenuniv.nl/pub/ldac/software/pipeline.pdf}})
C programs on catalogs extracted from reduced pixel data.  The C programs are
wrapped in \py\ to allow interaction with the object-oriented database model
employed by \awe.  In local astrometry, all the steps in the astrometric
solution (pre-astrometric correction, association, formal solution, etc.) are
handled by the \texttt{LDAC} programs.  In global astrometry, all the steps are
also handled by \texttt{LDAC} except for the initial cross-correlation (called
association) of sources which is handled by the \awe\ database (via advanced
queries).  This offers a performance advantage because the data to be
associated already resides in the database to be used in any combination as
needed.

\subsubsection{Local astrometry}\label{sec:local}

Local astrometry in \awe\ starts with a \reduc\ that has some basic astrometry,
directly from the telescope or updated sometime prior ingestion.  In a parallel
environment, the \reduc\ is run through the {\astro}Task, a \py\ convenience
recipe interacting with the database, whereby various C programs wrapped in
\py\ solve for the astrometry on the catalog level.  \texttt{SExtractor} is
run to extract the initial catalog.  After this, LDAC tools perform all
subsequent operations: pre-astrometric fitting to solve for large
(approximately arcminute level) offsets, scaling, and rotations using the any
all-sky catalog for reference (e.g., USNO, 2MASS, etc.).  This pre-astrometry
is then applied to the catalog and it is formally associated with the reference
catalog with offsets that are now on the order of arcseconds.  During the
process, only the most stellar-like and best quality objects, as determined by
SExtractor flags (for saturation, incomplete objects on the edge of a detector
chip, blended objects, etc.) are retained.  The catalog is then run through the
\texttt{LDAC.astrom} program where the final astrometry is determined
(least-squares fit to a 2-degree polynomial) and residuals catalog created.
The last step is converting the distortion correction to world coordinates
prior to storing the solution parameters in the database and the residuals
catalog on the dataserver.  These final residuals are now on the level of
accuracy of the reference catalog used.

The residuals catalog output from the \texttt{LDAC.astrom} program contains
residuals of the form DRA $=$ RA$_{ref} - $RA$_{ldac}$ and DDEC $=$ 
DEC$_{ref} - $DEC$_{ldac}$, where RA$_{ldac}$ and DEC$_{ldac}$ are the
coordinates of the extracted sources, corrected for all distortions by the LDAC
programs, and RA$_{ref}$ and DEC$_{ref}$ are the coordinates of the reference
sources from the reference catalog used.  The residual plots created by the
\astro\ \texttt{inspect} method plot information directly from this residuals
catalog and show what is to be the expected precision of the correction when
the \reduc\ is regridded into a \regri.  After the local astrometric solution
is created, the information can be applied to create a \regri\ (see
Sect.~\ref{sec:regri}) and eventually a \coadd\ (see Sect.~\ref{sec:coadd}).

\begin{figure}
\begin{center}
\includegraphics[angle=0,width=94mm]{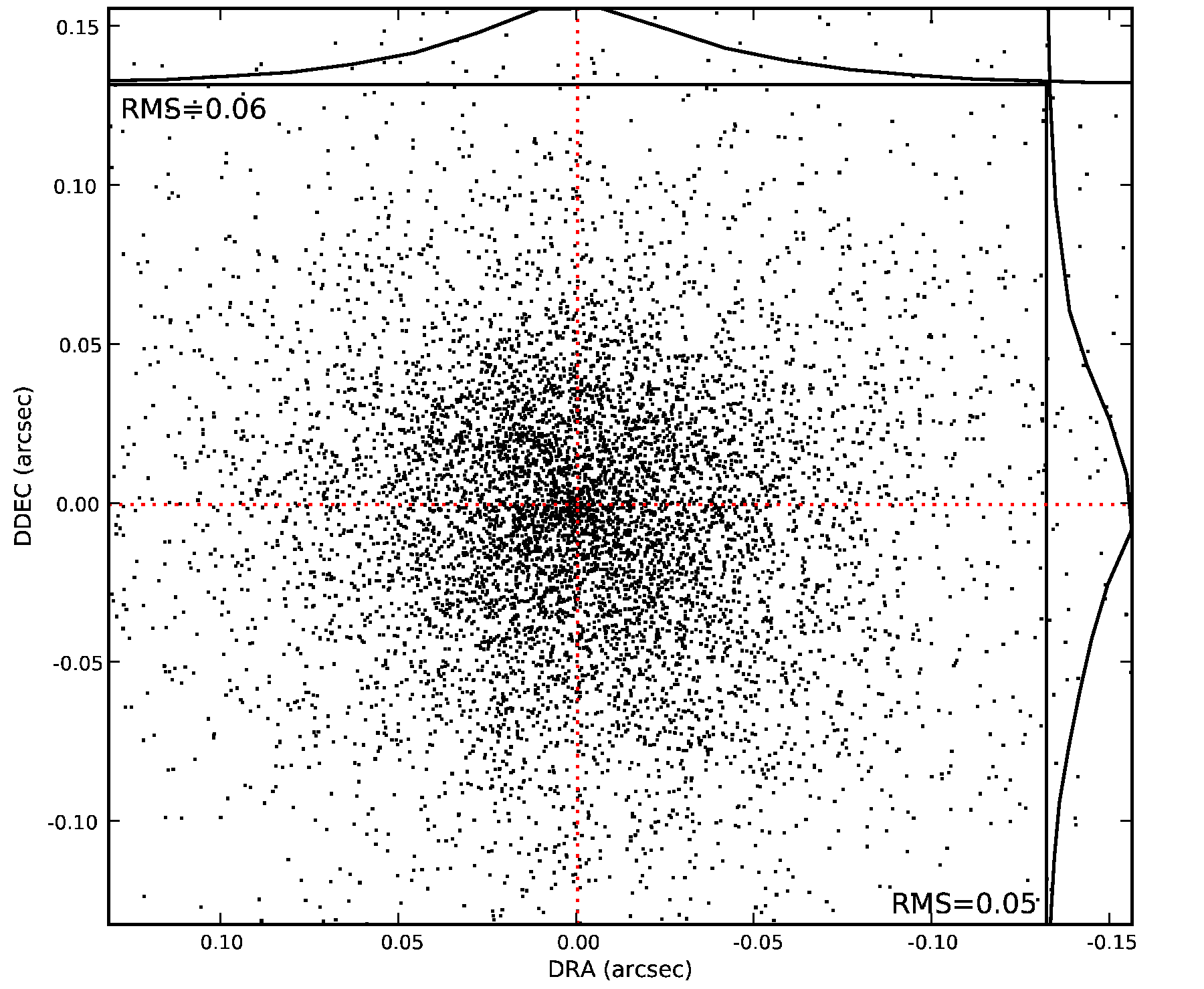}
\includegraphics[angle=0,width=94mm]{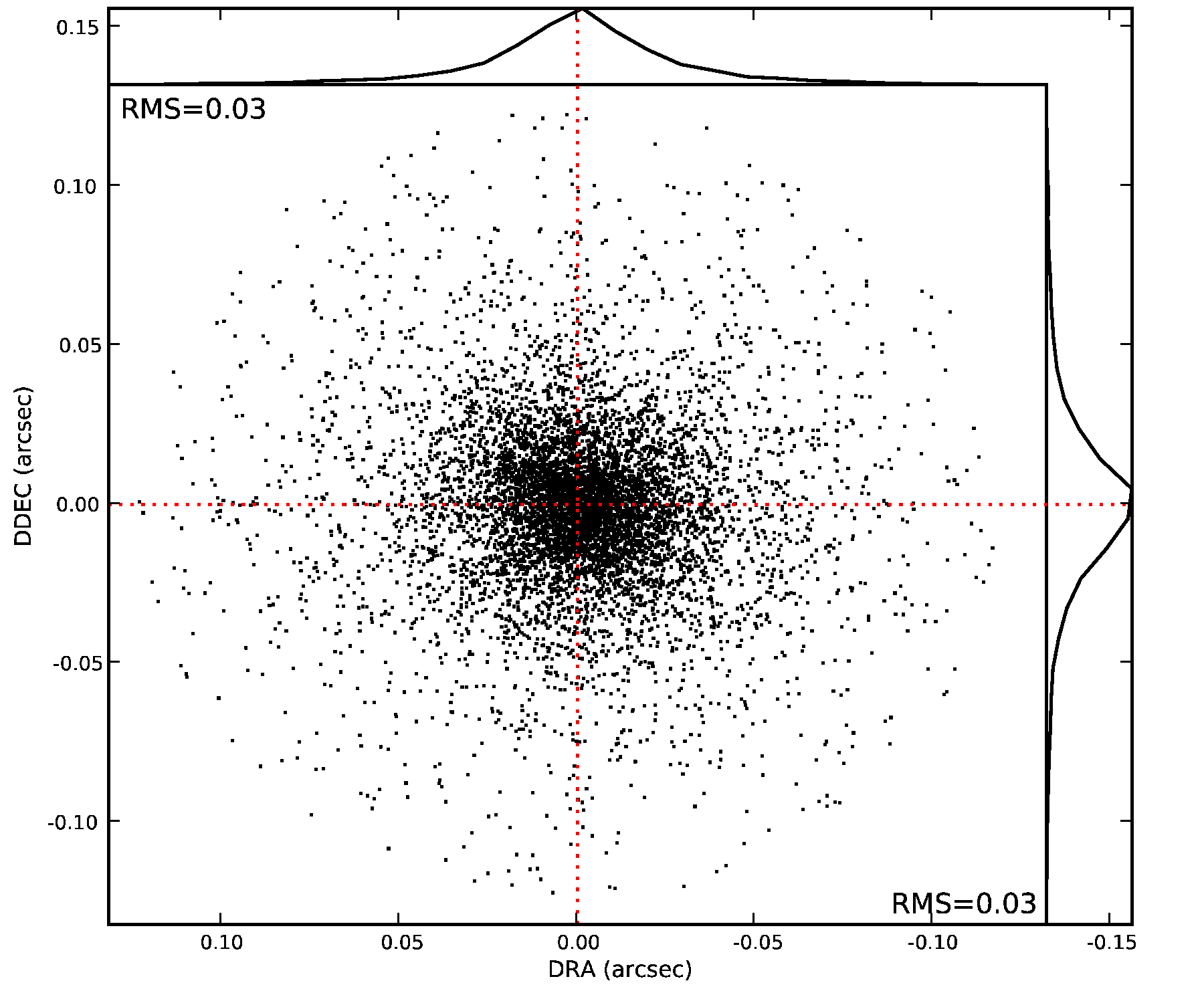}
\caption{
An example of improvement from the local to the global solution.  Both panels
show astrometric residuals, in arcsec, $\Delta RA = RA_1 - RA_2$ and $\Delta
Dec = Dec_1 - Dec_2$, where $RA_1$ and $DEC_1$ are the source positions from
any one frame, and $RA_2$ and $DEC_2$ are the source positions of all matching
sources in one of the other frames, same or different detector chip, that
overlaps the first.  The top panel shows the overlapping source position
differences from 32 frames of a 4-point WFI dither regridded using the
\textit{local solution} (limits scaled to match lower panel), the bottom panel
shows the same for the same frames regridded using the \textit{global
solution}.  The improvement in precision is greater than a factor two.
}\label{fig:astro}
\end{center}
\end{figure}

\subsubsection{Global astrometry}\label{sec:globa}

The most important concept in the global solution in \awe\ is that it is \textit{local}.  It is local in the sense that it uses the extra information of a set
of dithered observations that are closely matched both temporally and spatially
(e.g., exposures taken within one to two hours with more than 90\% of each
detector chip participating in the overlap region, respectively)\footnote{
Global astrometry in \awe\ is based on the concept of fixed focal-plane
geometry.  This means that any difference in the \textit{apparent} focal plane
from pointing to pointing is assumed to change in a linear fashion only, with
higher order distortions remaining constant (i.e., only relative translations
of the entire focal plane in RA and Dec are corrected for).  This asumption of
fixed focal-plane geometry adds information to the system, benefiting the
astrometric solution.  Generally, only sets of exposures taken temporally close
and spatially close will match this criteria.  These two conditions minimize
differences in telescope flexure caused by different altitude and azimuth
locations, and maximize the number of objects common to all exposures.
}.  The extra information characteristic of a closely matched dither consists
of the smooth variations in time of the optical system distortions and the
large amount of overlap of the detector area.  Combining the distortion
information with the overlap information allows the global solution to attain
the higher precision needed for proper coaddition of the source frames.  This
\textit{local}-global astrometry is the only method of global astrometry
certified in \awe.

The process of \textit{global}-global astrometry is quite different.  It
involves combining those dithers from widely different observation times, using
independent derivations of the optical system distortions, but combining all
overlap information available from overlapping dithers.  It allows for the
discontinuity among dithers that the \textit{local}-global process cannot.
This type of global astrometry is not present in \awe\ at this time.

Global astrometry in \awe\ starts with the \texttt{GAstromSourceListTask}, a
\py\ convenience recipe interacting with the database that creates a special
\sourc\ (see Sect.~\ref{sec:sourc}) from the source \reduc\ using the \astro\
information created by the local solution.  This is done in a parallel
environment and only if the {\sourc}s don't already exist.  Next, the
\texttt{GAstromTask} recipe is run in a serial environment as only a single
thread is needed.  It associates the source position information from the
associated \sourc, residing solely in the database, using an \assoc\ object
(see Sect.~\ref{sec:assoc}).  This step replaces the \texttt{LDAC.associate} stage
in the local solution.  After the association, \texttt{LDAC.astrom} is run on
the associated data using a least-squares fit to a 3-degree polynomial (as
opposed to a 2-degree polynomial in the local solution), and like the local
solution, a residuals catalog is created.

The residuals catalog output from the \texttt{LDAC.astrom} program in this case,
contains two sets of residuals, one identical to that of the local solution
with respect to the reference catalog used (see Sect.~\ref{sec:local}), and
the other with respect to the overlapping extracted sources.  The latter
residuals are of the form DRA $=$ RA$_2 - $RA$_1$ and DDEC $=$ DEC$_2 -
$DEC$_1$, where RA$_1$ and DEC$_1$ are the coordinates of the extracted sources
from a given frame and RA$_2$ and DEC$_2$ are the coordinates of the extracted
sources from another pointing, same or different detector chip, that overlaps
the first, both corrected for all distortions by the \texttt{LDAC.astrom}
program.  The residual plots created by the \gastr\ \texttt{inspect} method
plots both sets of residuals directly from this residuals catalog, both by
individual detector chip and for all detector chips combined, and shows what is
to be the expected precision of the global solution used to combine a set of
{\regri}s into a \coadd.

After the global astrometric solution is created, the information is used to
create a new \astro\ instance for each \reduc\ that went into the solution.
The parameters and statistics for the global solution are computed and stored
on a per frame basis and likely will not match those values of other frames
from the same solution.  As with the local solution, these parameters can be
applied to create {\regri}s (see Sect.~\ref{sec:regri}) and eventually a \coadd\
(see Sect.~\ref{sec:coadd}), but with much greater precision than with the local
solution only (see Fig.~\ref{fig:astro} for an example using WFI data).

\subsection{\photo}\label{sec:photo}

The photometric pipeline in \awe\ is aimed at calibrating large imaging surveys
taken with multi-detector chip wide-field imagers during many nights and
different epochs. Instrumental characteristics specific for wide-field imagers
need to be accounted for in a survey photometric pipeline. For example:
\begin{itemize}
\item
detector chip-to-detector chip variations. Each detector chip has its own small
and large-scale variations in pixel gain and can have a different median gain.
There can also be detector chip-to-detector chip variations in the
non-linearity behavior of count rates or color terms of the photometric
calibration.
\item
Illumination variation. Several wide-field imagers are known to have
illumination variations (e.g.,  MEGACAM at CFHT, WFI at ESO/MPG 2.2.m). The
gain variation over individual detector chips is characterized by flatfields
under the assumption of an ideal flat illumination over the field of view. In
practice this ideal flat illumination can be affected by stray and/or scattered
light (sky concentration) yielding variations of up to a few tenths of a
magnitude in amplitude.
\item
Shutter timing. The large FoV requires carefully designed shutter mechanisms.
Shutter timing variations might result in position dependent exposure times.
\end{itemize}

Performing a survey with such an instrument poses several challenges for the
photometric calibration. Long term, short-term, night-to-night or even
intra-night variations need to be monitored to create a homogeneous photometric
calibration across the whole survey area and survey time. It might be the case
that the very precise photometric calibration is dependent on instrumental
variations not captured by a single or handful standard star observations per
night, e.g., telescope altitude and azimuth. To detect and quantify all such effects
it is important to explore trends in photometric results as a function of many
parameters. To obtain the maximum photometric accuracy it is required to have
observations of photometric standards that densely cover the full FoV.

The goal of \awe\ photometric calibration is to establish the photometric
system resulting from the signal progressing through Earth's atmosphere,
telescope, filter, wide-field camera and each detector chip resulting in a
digital read-out. The photometric system is characterized in \awe\ in terms of
a multiplication of gains:
\begin{eqnarray}\label{eqn:gains}
\nonumber I_{obs} = g_{ff}(t,N,X,(x,y)) \times
                    g_e(t_0)g_e(t) g_{sel.e}(X)  \times \\
                    g_{qe}(t_0,N,X)g_{qe}(t,N,X) \times
                    g_{illum}(t,N,(x,y),X) \times
                    I_{ref},
\end{eqnarray}
where $I_{obs}$ is the observed countrate of a standard star in digital units
and $I_{ref}$ its emitted physical flux, $t$ is time and $(x,y)$ position on
the detector chip. The gain $g_{ff}(t,N,X,(x,y))$ characterizes the flat field.
The gains $g_e$ characterize the atmospheric extinction: $g_e({t_0})$ is the
scaling at time of the selected atmospheric extinction curve $g_{sel.e}(X)$
that is a function of filter $X$ and $g_{e}(t)$ models the change at time $t$.
The gains $g_{qe}$ characterize the overall instrumental quantum efficiency
that includes the light losses through the optics and conversion from physical
units of flux to countrates for detector chip $N$. The illumination variation
is captured as a separate gain $g_{illum}$.

By determining the gains, \awe\ then gives for each detector chip independently
the photometric calibration at any time for any pixel for each filter. The
photometric calibration objects have timestamps to indicate their validity
range in time (see Sect.~\ref{sec:conte}). Thus \awe\ holds a continuous
representation of the photometric system of an instrument if the calibration
plan of the instrument provides the required observations. This is another
example of how \awe\ calibrates the instrument instead of a specific data set.

The gain factors representing atmospheric extinction and instrumental quantum
efficiency are solved in magnitude space. The involved physics for wide-field
cameras is well-represented by the common photometric equation for astronomical
imagers:
\begin{eqnarray}\label{eqn:photo}
\nonumber m_{inst} = -2.5\log(countrate) + ZPT - k \times AM + \\
              C_0 - C_1 \times (m_{X2} - m_{X3})
\end{eqnarray}
where $m_{inst}$ is the magnitude of the object in the instrumental photometric
system, the $countrate$ is in ADU/s, k is the atmospheric extinction
coefficient, AM is the airmass, and $C_{0,1}$ are the terms describing the
corrections to go from the standard photometric system to the instrumental
photometric system. $m_{X2}-m_{X3}$ is the color between filter $X2$ and $X3$
of the standard star as listed in the catalog of the standard photometric
system.

\subsubsection{Atmospheric extinction}

The atmospheric extinction in magnitude space is assumed to be a linear
function of airmass AM (i.e.,$k \times AM$ in Eqn.~\ref{eqn:photo} which
is a representation of $g_e$ in Eqn.~\ref{eqn:gains} ($g_e \sim 10^{-(k
\times AM)/2.5}$). The task is to establish the atmospheric extinction
coefficient $k$. The airmass is taken from the observational metadata.

In \awe, the correction for the atmosphere in the photometric calibration can
be derived in four ways.
\begin{enumerate}
\item
Using a pre-defined atmospheric extinction coefficient.  The coefficient is
multiplied by the airmass (see Eqn.~\ref{eqn:photo}). These are stored in
the \awe\ database for each combination of instrument and filter object of the
class \texttt{Atmos\-pheric\-Extinction\-Coefficient}.
Users can insert their own atmospheric extinction coefficients in \awe.
\item
Using a pre-defined atmospheric extinction coefficient plus a shift. It is
using the coefficient just described plus a shift given by a report represented
by the class \texttt{Photometric\-Extinction\-Report}.
This kind of atmospheric correction on the photometry is represented by the
class \texttt{Atmospheric\-Extinction\-Curve}.
\item
Using standard star field observations. This kind of correction is represented
by the class \texttt{Atmospheric\-Extinction}. There are two sub options here:
\begin{enumerate}
\item
Using a single standard star field observation and a given zeropoint.
Eqn.~\ref{eqn:photo} is the used to determine an atmospheric extinction
coefficient. This type of atmospheric correction is represented by the
\texttt{Atmospheric\-Extinction\-Zeropoint} class.
\item
Using two observations of standard star fields at different airmass. By
equalling the zeropoint in Eqn.~\ref{eqn:photo} one can solve for the
atmospheric extinction coefficient. This correction type is represented by the
class \texttt{Atmospheric\-Extinction\-Frames}.
\end{enumerate}
\end{enumerate}

\subsubsection{Color terms}

Differences in the effective throughput per wavelength of the photometric
system of the standard system and the instrumental system can be caused, for
example, by differences in filter transmission curves or in quantum efficiencies
of the detector chips. In \awe, it is assumed that these differences can be
captured by a linear function of the standard star color in the standard
photometric system:
\begin{eqnarray}
m_{ref,i,inst}=m_{ref,i,std}+C_0 -C_1 \times (m_{i,X2}-m_{i,X3}),
\end{eqnarray}
where $m_{ref,i,inst}$ is the magnitude of the standard star $i$ in the
instrumental photometric system and $m_{ref,i,std}$ in the standard photometric
system.  For each combination of instrument and filter the two coefficients are
pre-determined and stored in \awe. The \texttt{Phot\-Transformation} class
represents the color transformation, and objects of this class contain the
coefficients. The magnitudes of the standard star in filters $X2,X3$ is taken
from the standard star catalog (a \texttt{Phot\-Ref\-Catalog} object) stored in
\awe.

\subsubsection{Zeropoints}\label{sec:zerop}

The flux counts and astrometry of stars in a photometric standard field are
measured using SExtractor.  The resulting catalog is associated (using the
prephotom package in LDAC) with known standard stars listed in a reference
catalog. Now a ``raw'' instrumental magnitude ($m_{raw,i,inst}$) and zeropoint
$ZPT_{raw,i,inst}$ are computed for each observed standard star $i$:
\begin{eqnarray}
m_{raw,i,inst} = -2.5 \log countrate \\
ZPT_{raw,i,inst}= m_{ref,i,inst} - m_{raw,i,inst}
\end{eqnarray}
A clipping is applied on the set of raw zeropoints:
\begin{equation}
|ZPT_{raw,i,inst} - median({ZPT_{inst,i,raw}})| < MAX\_MAG\_DIFF,
\end{equation}
with $MAX\_MAG\_DIFF$ set by the user. The result is stored in a photometric
source catalog represented by the class \phots.

If at least a required minimum number of standard stars identified in the
observation remain (the MIN\_NMBR\_OF\_STARS parameter), the final zeropoint is
computed. A sigma clipping with a threshold factor SIGCLIP\_LEVEL set by the
user is applied once to the raw zeropoints. The variance weighted mean and its
uncertainty are computed from the remaining raw zeropoints. This mean is then
corrected for the atmospheric extinction yielding the zeropoint $ZPT$. The
$ZPT$ is stored in a \photo\ object.  Formal errors are
propagated from count measurements through the computation of zeropoint and
atmospheric extinction.

\awe\ contains a photometric reference catalog that contains the
magnitudes of standard stars in Johnson-Cousins system (from Landolt and
Stetson), and the Sloan system in 22 SA fields.  By default, all entries are
used from in the standard star catalog, but one can limit this to subsets. It
is also possible to use a custom photometric reference catalog.

\subsubsection{\texttt{IlluminationCorrection}}\label{sec:illum}

The photometric calibration accounts for gain variations under the assumption
of an ideal flat illumination over the field of view.  In practice this ideal
flat illumination can be affected by stray light (sky concentration) and a
correction for this effect has to be made.

It is assumed that the effect of the illumination variations is larger than
detector chip-to-detector chip systematic variations. It has been verified that
this holds for the MEGACAM and WFI instruments. The starting point is all detectors of a mosaic of a
standard star field observation that is detrended up to the flat field level in a
given filter.  The raw zeropoint (see Sect.~\ref{sec:zerop}) is determined for
each standard star.  The residual between these zeropoints and their median
value over all detector chips in the mosaic is a measure of the illumination
variation.  The residual distribution is assumed to be well-fitted with a
two-dimensional second order polynomial (as is verified for MEGACAM and WFI)
using a chi-square minimization.  An illumination variation frame is created
from the polynomial fit for each detector chip. Each standard star field frame
is divided by this \illum\ and a new zeropoint determination is performed per
detector chip.  This last step corrects for any remaining detector
chip-to-detector chip variations.

The resulting illumination correction is applied to {\reduc}s in the following
manner: the background is removed from the science frames and the remaining
pixels associated with sources (both calculated by SExtractor) are multiplied
by the \illum.  The background is added back and the zeropoints from the
standard star field with illumination correction are applied.

In wide-field instruments (e.g., OmegaCAM), the illumination variation pattern
across the large detector block can vary with time, telescope position, etc.
In these cases, an \illum\ may fail to properly characterize the illumination
variation and require a different approach.  One such approach involves
compensating for only the pixel-to-pixel variations in the flat-fielding as
alluded to in Sect.~\ref{sec:maste}.  A \maste\ constructed from only the high
spatial frequencies of a \domef\ can be used to eliminate the pixel-to-pixel
sensitivity variations without adding any illumination variation from the low
spatial frequency (large scale) contributions.  Any remaining illumination
variation above the background, if it exists, can then be corrected for
appropriately, either as described above or via robust sky subtraction
techniques (e.g., with SExtractor).

\section{Image Pipeline: combining the pixels}\label{sec:image}

\begin{figure}
\begin{center}
\includegraphics[angle=0,width=118mm]{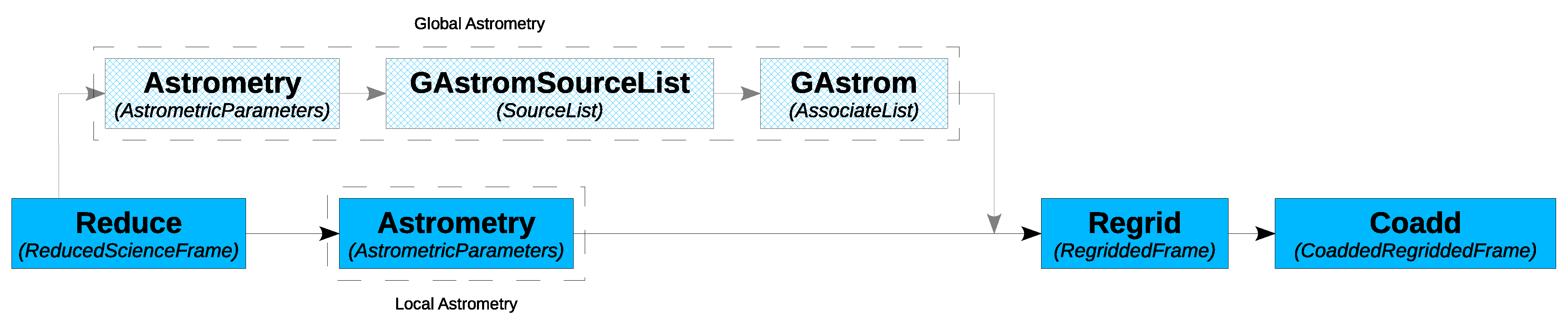}
\caption{
Schematic flow of the image pipeline following the coloring in Fig.~\ref{fig:targe}.  The recipes, also called \texttt{Tasks}, used to produce
various {\proce}s are indicated in each box (with their data product in
parentheses) and described in the various sections.  The arrows connecting them
indicate the direction of processing.  Note that the global (multi-chip)
astrometry branch is optional and supplementary to the local (single-chip)
astrometry.  Also note, that while \assoc\ is the formal data product of
\texttt{GAstrom}, new \astro\ objects are created in the process as well.
}\label{fig:image}
\end{center}
\end{figure}

As mentioned earlier, one advantage of the \awe\ is its parallel processing
capability.  Much of the processing is done in a parallel environment, one
detector chip per CPU node.  There are two places in the image pipeline,
however, where the information of individual detector chips must be combined:
the astrometric solution may be derived for all detector chips simultaneously
(\textit{global astrometry}), and science images may be coadded into larger
mosaics and/or deeper images.  See Fig.~\ref{fig:image} for an overview.

Many \proce's have configurable \textit{processing parameters} to control how
they are processed.  Table~\ref{tab:image} gives an overview of these
\texttt{process\_params} for the image pipeline.

\begin{table}
\begin{center}
\begin{tabular}{|l|l|r|r|}
\hline
\textbf{Class} & \textbf{process\_param}     & \textbf{value} & \textbf{units}\\
\hline
\reduc & overscan\_correction                &           6    &           \\
       & fringe\_threshold\_low              &           1.5  &           \\
       & fringe\_threshold\_high             &           5.0  &           \\
       & image\_threshold                    &           5.0  &           \\
\hline
\satur & threshold\_low                      &          50.0  &       ADU \\
       & threshold\_high                     &       50000.0  &       ADU \\
\hline
\satel & detection\_threshold                &           5.0  &           \\
       & hough\_threshold                    &        1000.0  &           \\
\hline
\regri & background\_subtraction\_type       &           0    &           \\
\hline
\sourc & htm\_depth                          &          25    &           \\
\hline
\assoc & search\_distance                    &           5.0  &    arcsec \\
       & single\_out\_closest\_pairs         &           1    &           \\
       & sextractor\_flag\_mask              &           0    &           \\
\hline
\end{tabular}
\end{center}
\caption{
Processing parameters and their default values.  These values are
representative of the typical value for any instrument.  Some instruments may
have values that different from these based on experience with that instrument.
See the document page linked from the class name or appropriate links on
\url{http://doc.astro-wise.org/astro.main.html} for more details.
}\label{tab:image}
\end{table}

\subsection{\reduc}\label{sec:reduc}

The most basic outcome of the image pipeline is the \reduc.  Conventional
de-trending steps are performed when making this frame:

\begin{enumerate}
\item
overscan correction and trimming
\item
subtraction of the \biasf
\item
division by the \maste
\item
scaling and subtraction of a \fring\ if indicated
\item
multiplication by an \illum\ if indicated
\item
creation of the individual weight image
\item
computation of the image statistics
\end{enumerate}
Please note that:
\begin{itemize}
\item
the overscan correction can be a null correction (i.e., no modification
of the pixel values)
\item
the illumination correction step (i.e., application of a photometric flat field)
has had a SExtractor-created background removed and then reapplied after the
multiplication, and the correction only occurs when requested and if a suitable
\illum\ exists
\end{itemize}

\subsection{\weigh}

In addition to the effects of hot and cold pixels, individual images may be
contaminated by saturated pixels, cosmic ray events, and satellite tracks.  For
purposes of subsequent analysis and image combination, affected pixels unique
to each image need to be assigned a weight of zero in that image's weight map.

Since the variance is inversely proportional to the Gain, which is
proportional to the flatfield, the weight is given by:
$$W_{ij} = G_{ij}\,P_{hot}\,P_{cold}\,P_{saturated}\,P_{cosmic}\,P_{satellite},$$
where $W_{ij}$ is the weight of a given pixel, $G_{ij}$ is the gain of a given
pixel (taken from the flat field), and the rest of the members are binary maps where
good pixels have a value of 1 and bad pixels have a value of 0.  These maps
are, respectively, a \hotpi, a \coldp, a \satur, a \cosmi, and a \satel, the
last three being calculated directly from the \reduc\ after detrending.

\subsubsection{\satur}
Saturated pixels are pixels whose counts exceed a certain threshold.  In
addition, saturation of a pixel may lead to \textit{dead} neighbouring pixels,
whose counts lie below a lower threshold.  These upper and lower thresholds are
defined and stored in the object.

\subsubsection{\cosmi}

Two programs may be used to detect cosmic ray events:
\begin{enumerate}
\item
\textbf{SExtractor} can be run with a special filter that is only sensitive to
cosmic-ray-like signal. This requires a `retina' filter, which is a neural
network that uses the relative signal in neighboring pixels to decide if a
pixel is a cosmic.  A retina filter, called 'cosmic.ret' is provided.  Run
SExtractor with \texttt{FILTER\_NAME=cosmic.ret}, to run SExtractor in comic
ray detection mode.  This results in a so-called segmentation map,
recording the pixels affected by cosmic ray events.  This segmentation can be
used to assign a weight of zero to these pixels.
\item
\textbf{CosmicFITS} is designed as a stand-alone program to detect cosmic ray
events.
\end{enumerate}
In the \awe, the SExtractor method is the preferred cosmic ray event detection
method.

\subsubsection{\satel}

Linear features can be detected using a \textit{Hough transform} algorithm,
which is used to find satellite tracks.  See \citet{hough,duda} for
more information about the Hough transform.

A point $(x, y)$ defines a curve in Hough space $(\rho, \theta)$, where:
$$\rho = x\,cos \theta + y\,sin \theta,$$
corresponding to lines with slopes $0<\theta<\pi$, passing at a distance $\rho$
from the origin.  This means that different points lying on a straight line in
image space, will correspond to a single point ($\rho, \theta$) in Hough space.

The algorithm then creates a Hough image from an input image, by adding a
Hough curve for each input pixel which lies above a given threshold.  This
Hough image (effectively a histogram of pixels corresponding to possible lines)
is clipped, and transformed back into a pixelmap, masking lines with too many
contributing pixels.

\subsection{\astro}

The parameters from the astrometric solution are used during the regridding
process and their creation has already been discussed in Sect.~\ref{sec:astro}.

\subsection{\photo}

The parameters from the photometric solution are used during the coaddition
process and their creation has already been discussed in Sect.~\ref{sec:photo}.

\subsection{\regri}\label{sec:regri}

Regridding and co-adding are done using the
SWarp\footnote{\url{http://astromatic.iap.fr/software/swarp/}} program.  Before
images are co-added, they are resampled to a predefined pixel grid (see
Sect.~\ref{sec:skygr}).  By co-adding onto a simple coordinate system,
characterized by the projection (Tangential, Conic-Equal-Area), reference
coordinates, reference pixel, and pixel scale, the distortions recorded by the
astrometric solution are removed from the images.  To this end a set of
projection centers is defined, at 1 degree separation and pixel scale of 0.2
arcsec.  A \reduc\ resampled to this grid is called a \regri.  The background
of the image can be calculated and subtracted at this time, if desired.

\subsection{\coadd}\label{sec:coadd}

After the {\regri}s are made, it is only a matter of applying the photometry of each frame and stacking the result.  This process creates a \coadd.

One point of great importance in considering the coadded data is its pixel
units.  The units are fluxes relative to the flux corresponding to magnitude=0.
In other words, the magnitude $m$ corresponding to a pixel value $f_0$ is:
\begin{equation}
\label{coaddunit}
m = -2.5\,log_{10} f_0
\end{equation}

The value $f_{out}$ of a pixel in the \coadd\ is computed from all overlapping
pixels \textit{i} in the input {\regri}s according to this formula:
\begin{equation}
f_{out}=\Sigma_i(w_i*FLXSCALE_i*f_i) / \Sigma_i(w_i),
\end{equation}
where $f_i$ is the pixel value in the \regri, $FLXSCALE_i$ is calculated from
the zeropoint, and $w_i=weight_i/FLXSCALE_i^2$ where $weight_i$ is the value of
the pixel in the input weight image.  A \weigh\ is created as well. The
value $w_{out}$ of the pixel in the weight frame for the coadd is:
\begin{equation}
w_{out}=\Sigma_i(w_i)
\end{equation}

\subsection{\sourc}\label{sec:sourc}

In \awe, source information from processed frames can be stored in the database
in the form of {\sourc}s.  These are simply a transcription of a
SExtractor-derived catalog values (position, ellipticity, brightness, etc.) into
the database.  Normally, the catalog was derived from a processed frame
existing in the system, but this is not a requirement.  Arbitrary SExtractor
catalogs meeting a minimum content criteria can be ingested as well.  This is
how large survey results and reference catalogs are brought into the system.

These {\sourc}s can be used for a variety of purposes such as astrometric and
photometric correction, but are normally an end product of the image pipeline
storing key quantities about the sources in question for further analysis.
Multiple {\sourc}s can be combined into an \assoc, and later into another
\sourc\ via the \combi\ machinery.

\subsection{\assoc}\label{sec:assoc}

Multiple {\sourc}s can be spatially combined (VIA RA and DEC values) and stored
in the database via the \assoc\ class.  The association is done in the
following way:
\begin{enumerate}
\item
The area of overlap of the two {\sourc}s is calculated.  If there is no overlap
no associating will be done.
\item
The sources in one \sourc\ are paired with sources in the other if they are
within a certain association radius.  Default radius is 5$\prime\prime$.  The
pairs get an unique associate ID (AID) and are stored in the \assoc. A filter
is used to select only the closest pairs.
\item
Finally the sources which are not paired with sources in the other list and are
inside the overlapping area of the two \sourc\ are stored in the \assoc\ as
singles.  They too get an unique AID.
\end{enumerate}

Very important is the type of association being done.  One of three types:
chain, master or matched, will be done.  In a \textit{chain} association, all
subsequent {\sourc}s are matched to the previous \sourc\ to find pairs, in a
\textit{master} association, they are always matched with the first \sourc, and
in a \textit{matched} association, all {\sourc}s are matched with all other
{\sourc}s.

\section{Summary}\label{sec:summa}

The development and implementation of the \aw\ optical pipeline has been
described.  This pipeline uses the \awenv: an information system designed to
integrate hardware, software and human resources, data processing, and quality
control in a coherent system that provides an unparalleled environment for
processing astronomical data at any level, be it an individual user or a large
survey team spread over many institutes and/or countries.

The \awenv\ is built around an Object-Oriented Programming (OOP) model using
\py where each data product is represented by the instantiation of a particular
type of object.  The processability and quality of these data objects
({\proce}s) is moderated by built-in attributes and methods that know, for each
individual type of object or OOP class, how to process or qualify itself.  All
progenitor and derived data products are transparently linked via the database,
providing an uninterrupted path between completely raw and fully processed
data.

This data lineage and provenance allows for a type of processing whereby the
pipeline used for a given set of data is created \textit{on-the-fly} for that
particular set of data, where the Unix \texttt{make} metaphor is employed to
chain backward though the data, processing only what needs to be processed (target processing).
This allows unparalleled efficiency and data transparency for reprocessing the
data when necessary, as the raw data is always available when newer techniques
become available.

Calibration of data follows the usual routes, but has been optimized for
processing of OmegaCAM calibration data meant for detrending survey data.  In
this process, data is processed and reprocessed as more and more knowledge of
the instrument system (from the optics through detector chain) is acquired.
This effectively calibrates the instrument, leaving the data simply to be
processed without the need of users find or qualify their own calibrations.
Various attributes of calibration objects (validity, quality, valid time
ranges) transparently determine which calibrations are best to be used for any
data.  Processing parameters are set and can be reset as desired.  These
parameters are retained as part of the calibration object and guarantee that a
given object can be reprocessed to obtain the same result or be
\textit{tweaked} to improve the result.  The processing of science data is
governed by the same validity, quality, valid time range, and processing
parameter mechanism that is used for calibration data.

The calibration pipeline starts with a \readn\ object created from {\rawbi}s
that is used to determine a clipping limit for \biasf\ creation.  A \gainl\
object can be processed from a special set of {\rawdo}s taken for the purpose.
From this result, both the gain (in $e^-/$ADU) and the detector linearity can
be determined.  A master \biasf\ is created from a set of {\rawbi}s to remove
2-dimensional additive structure in detectors.  The \darkc\ is measured for
quality control of the detectors, but is not applied to the pixels.  Bad pixels
in a given detector can be found from the \biasf\ and a flat field image.
These are termed \hotpi\ and \coldp, respectively.

Flat field creation in \aw\ can be very simple or very complex.  On the simple
side, a single set of {\rawdo}s or {\rawtw}s can be combined with outlier
rejection and normalized to the median.  On the complex side, high spatial
frequencies can be taken from the \domef and the low spatial frequencies from
the \twili.  In addition, a \night\ can be added to improve this result.  For
an additional refinement to the flat field correction for redder filters, a
\fring\ can be created.

Astrometric calibration starts with extraction of sources from individual
{\reduc}s.  The source positions are matched to those in an astrometric
reference catalog (e.g., USNO-A2.0) and all the positional differences
minimized with the LDAC programs.  This \textit{local} solution can then be
further refined by adding overlap information from a dither to form a
\textit{global} astrometric solution.  Astrometric solutions are always stored
for each \reduc\ individually.  Photometric calibration also starts with source
extraction (as a \phots) and positional association.  Then, the magnitudes of
the associated sources are compared to those in a photometric reference catalog
(e.g., Landolt) and the mean of the Kappa-sigma-clipped values results in a zeropoint
for a given detector for the night in question.  The extinction can be derived
from multiple such measurements, the results of both being stored in a \photo\
object.  As an optional refinement to the photometric zeropoint, a photometric
super flat can be constructed by fitting magnitude differences as a function of
radius across the whole detector block.  The result of this is stored in an
\illum\ object.

The image pipeline takes all the calibrations from \biasf\ through \maste\ to
transform a \rawsc\ into a \reduc.  This includes trimming the image after
applying the overscan correction, subtracting the \biasf, dividing by the
\maste, and applying the \fring\ and \illum\ if necessary.  The \weigh\ is
constructed by taking the \hotpi\ and \coldp\ and combining them with a \satur,
a \satel, a \cosmi, and optionally a \illum.  These are all applied to the
\maste\ to create the final \weigh.  Next, the \astro\ is applied to the
\reduc\ in creating the \regri, and the \photo\ is applied to multiple
{\regri}s to form a \coadd.  Lastly, the sources from one \coadd\ can be
extracted into a \sourc and associated with other {\sourc}s to form an \assoc\
object.  This last is the final output of the image pipeline and can combine
information from multiple filters on the same part of the sky into one data
product.

Using \awe, The \kids\ survey team has begun processing each week's worth of
data taken at the VST (more than half a terabyte) in a single night.  The part
of the data that requires it (bad quality or validity) is reprocessed nightly
as necessary to gain the required insight into the different aspects of the
calibration process: detrending calibrations, astrometric calibrations, and
photometric calibrations.

The \awenv\ is a unique multi-purpose pipeline for astronomical surveys.  All
required tools (ingestion, processing, quality control, and publishing) are 
integrated in an intuitive and transparent way.  It has already been used to
process archive WFI@2.2m, MegaCam@CFHT (CFHTLS), and VIRCam@VISTA data in
pseudo-survey mode in preparation for its main task: processing \kids,
Vesuvio, OmegaWhite, and OmegaTrans survey data from the newly commissioned
OmegaCAM@VST.


\section{Appendix: skygrid of projection centers}\label{sec:skygr}

Tables \ref{tab:grid1} \& \ref{tab:grid2} describe a grid on the sky for
projection and co-addition purposes in a condensed format. It contains 95
strips as function of decreasing declination
($0^\circ\geq\delta\geq-90^\circ$). For each strip the size in degrees and the
number of $1^\circ\times1^\circ$ fields per strip is given.  The last column
contains the overlap between fields in \%. By mirroring the grid along the
equator one obtains a grid for the northern hemisphere. The combination of the
grids for both hemispheres is a grid for the entire sky.

\begin{table}[h!]
\begin{tabular}{lcccr}
\textbf{strip}&\textbf{$-\delta$ [$\circ$]}&\textbf{size [$\circ$]}&\textbf{fields/strip}&\textbf{overlap [\%]}\\
 1& 0.00&360.00&378&5.0\cr
 2& 0.96&359.95&378&5.0\cr
 3& 1.91&359.80&378&5.1\cr
 4& 2.87&359.55&378&5.1\cr
 5& 3.83&359.20&377&5.0\cr
 6& 4.79&358.74&376&4.8\cr
 7& 5.74&358.19&375&4.7\cr
 8& 6.70&357.54&374&4.6\cr
 9& 7.66&356.79&373&4.5\cr
10& 8.62&355.94&372&4.5\cr
11& 9.57&354.99&371&4.5\cr
12&10.53&353.94&370&4.5\cr
13&11.49&352.79&369&4.6\cr
14&12.45&351.54&368&4.7\cr
15&13.40&350.19&367&4.8\cr
16&14.36&348.75&366&4.9\cr
17&15.32&347.21&365&5.1\cr
18&16.28&345.57&363&5.0\cr
19&17.23&343.84&361&5.0\cr
20&18.19&342.01&359&5.0\cr
21&19.15&340.08&357&5.0\cr
22&20.11&338.06&355&5.0\cr
23&21.06&335.95&353&5.1\cr
24&22.02&333.74&350&4.9\cr
25&22.98&331.43&347&4.7\cr
26&23.94&329.04&344&4.5\cr
27&24.89&326.55&341&4.4\cr
28&25.85&323.97&338&4.3\cr
29&26.81&321.31&335&4.3\cr
30&27.77&318.55&332&4.2\cr
31&28.72&315.70&329&4.2\cr
32&29.68&312.77&326&4.2\cr
33&30.64&309.74&323&4.3\cr
34&31.60&306.64&320&4.4\cr
35&32.55&303.44&317&4.5\cr
36&33.51&300.16&314&4.6\cr
37&34.47&296.80&311&4.8\cr
38&35.43&293.35&308&5.0\cr
39&36.38&289.83&304&4.9\cr
40&37.34&286.22&300&4.8\cr
41&38.30&282.53&296&4.8\cr
42&39.26&278.76&292&4.7\cr
43&40.21&274.91&288&4.8\cr
44&41.17&270.99&284&4.8\cr
45&42.13&266.99&280&4.9\cr
46&43.09&262.92&276&5.0\cr
47&44.04&258.78&272&5.1\cr
48&45.00&254.56&267&4.9\cr
49&45.96&250.27&262&4.7\cr
50&46.91&245.91&257&4.5\cr
\end{tabular}
\caption{Strips 1-50}\label{tab:grid1}
\end{table}

\begin{table}[h!]
\begin{tabular}{lcccr}
\textbf{strip}&\textbf{$-\delta$ [$\circ$]}&\textbf{size [$\circ$]}&\textbf{fields/strip}&\textbf{overlap [\%]}\\
51&47.87&241.48&252& 4.4\cr
52&48.83&236.99&247& 4.2\cr
53&49.79&232.43&242& 4.1\cr
54&50.74&227.80&237& 4.0\cr
55&51.70&223.11&232& 4.0\cr
56&52.66&218.36&227& 4.0\cr
57&53.62&213.54&222& 4.0\cr
58&54.57&208.67&217& 4.0\cr
59&55.53&203.74&212& 4.1\cr
60&56.49&198.75&207& 4.1\cr
61&57.45&193.71&202& 4.3\cr
62&58.40&188.61&197& 4.4\cr
63&59.36&183.46&192& 4.7\cr
64&60.32&178.26&187& 4.9\cr
65&61.28&173.01&182& 5.2\cr
66&62.23&167.71&176& 4.9\cr
67&63.19&162.36&170& 4.7\cr
68&64.15&156.97&164& 4.5\cr
69&65.11&151.54&158& 4.3\cr
70&66.06&146.06&152& 4.1\cr
71&67.02&140.54&146& 3.9\cr
72&67.98&134.98&140& 3.7\cr
73&68.94&129.39&134& 3.6\cr
74&69.89&123.76&128& 3.4\cr
75&70.85&118.09&122& 3.3\cr
76&71.81&112.39&116& 3.2\cr
77&72.77&106.66&110& 3.1\cr
78&73.72&100.90&104& 3.1\cr
79&74.68& 95.11& 98& 3.0\cr
80&75.64& 89.30& 92& 3.0\cr
81&76.60& 83.46& 86& 3.0\cr
82&77.55& 77.59& 80& 3.1\cr
83&78.51& 71.71& 74& 3.2\cr
84&79.47& 65.80& 68& 3.3\cr
85&80.43& 59.88& 62& 3.5\cr
86&81.38& 53.94& 56& 3.8\cr
87&82.34& 47.98& 50& 4.2\cr
88&83.30& 42.01& 44& 4.7\cr
89&84.26& 36.03& 38& 5.5\cr
90&85.21& 30.04& 32& 6.5\cr
91&86.17& 24.05& 26& 8.1\cr
92&87.13& 18.04& 19& 5.3\cr
93&88.09& 12.03& 13& 8.1\cr
94&89.04&  6.02&  7&16.4\cr
95&89.90&  0.63&  1&   -\cr 
\end{tabular}
\caption{Strips 51-95}\label{tab:grid2}
\end{table}

\begin{acknowledgements}
Astro-WISE is an on-going project which started from a FP5 RTD programme funded
by the EC Action ``Enhancing Access to Research Infrastructures''. This work is
supported by FP7 specific programme ``Capacities - Optimising the use and
development of research infrastructures''.  Special thanks to Francisco Valdes
for his constructive comments.
\end{acknowledgements}

\end{document}